\documentclass[lettersize,journal]{IEEEtran}
\usepackage{amsmath,amsfonts}
\usepackage{algorithmic}
\usepackage{algorithm}
\usepackage{array}
\usepackage{textcomp}
\usepackage{stfloats}
\usepackage{url}
\usepackage{verbatim}
\usepackage{graphicx}
\usepackage{cite}
\usepackage{xspace}
\ifCLASSOPTIONcompsoc
\usepackage[caption=false, font=normalsize, labelfont=sf, textfont=sf]{subfig}
\else
\usepackage[caption=false, font=footnotesize]{subfig}

\begin{document}
\title{Planning and Tracking Control of Full Drive-by-Wire Electric Vehicles in Unstructured Scenario}
\author{Guoying Chen, Min Hua, Wei Liu, Jinhai Wang, Shunhui Song, Changsheng Liu
\thanks{Guoying Chen is with the State Key Laboratory of Automotive Simulation and Control, Jilin University, Changchun, 130022, P.R. China. (e-mail: cgy-011@163.com), Corresponding author.}
\thanks{Min Hua is with the School of Engineering, University of Birmingham, Birmingham, B15 2TT, UK. (e-mail: mxh623@student.bham.ac.uk).}
\thanks{Wei Liu is with the School of Electrical and Computer Engineering, Purdue University, West Lafayette, IN 47907, USA (e-mail: liu3044@purdue.edu).}
\thanks{Jinhai Wang is with the  School of Automotive Engineering, Wuhan University of Technology, Wuhan, 430070, P.R. China (e-mail: wangjinhai@whut.edu.cn).}
\thanks{Shunhui Song is with the School of Automotive Studies, Tongji University, Shanghai, 218074, P.R. China (e-mail: songshunhui@gmail.com).}
\thanks{Changsheng Liu is with 
the College of Computer Science and Technology, Zhejiang University, Hangzhou, 310027, P.R. China. (e-mail: changshengliu23@gmail.com).}}


\maketitle

\begin{abstract}
Full drive-by-wire electric vehicles (FDWEV) with X-by-wire technology can achieve independent driving, braking, and steering of each wheel, providing a good application platform for autonomous driving technology. Path planning and tracking control, in particular, are critical components of autonomous driving. However, It is challenging to comprehensively design an robust control algorithm by integrating vehicle path planning in a complicated unstructured scenario for FDWEV. To address the above issue, this paper first proposes the artificial potential field (APF) method for path planning in the prescribed park with different static obstacles to generate the reference path information, where speed planning is incorporated considering kinematics and dynamic constraints. Second, two tracking control methods, curvature calculation (CC-based) and model predictive control (MPC-based) methods with the lateral dynamics model, are proposed to track the desired path under different driving conditions, in which a forward-looking behavior model of the driver with variable preview distance is designed based on fuzzy control theory.  CarSim-AMESim-Simulink co-simulation is conducted with the existence of obstacles. The simulation results show that the proposed two control approaches are effective for many driving scenarios and the MPC-based path-tracking controller enhances dynamic tracking performance and ensures good maneuverability under high-dynamic driving conditions.

\end{abstract}


\section{Introduction}
\label{sec:introduction}
\IEEEPARstart{V}{ehicle} driving safety has always been an important research field for vehicle manufacturing \cite{guan2022discrete,  liu2018vehicle, chen2019dynamics, cheng2020robust}. With the increasingly intelligent, electrified, wire-controlled, and networked vehicles \cite{ xu2022opv2v}, the intelligent full drive-by-wire electric vehicle possesses more advantages than traditional vehicles, such as independent driving, braking, and steering control \cite{chen2019comprehensive, liu2021automated}. In addition, each wheel in the steering system can achieve positive and negative 180-degree rotation around the z-axis of the tire for various steering modes, e.g., crab, traverse, diagonal travel, and steer in place \cite{zhao2019identification}. Controllable degrees of freedom can improve the mobility and flexibility of the vehicle at low speeds to achieve automatic parking and other functions in low-speed and narrow spaces, as well as the stability and safety of the vehicle when driving at high speeds \cite{xia2018automated}. Furthermore, due to the application of X-by-wire technology, four-wheel motor torque, and other parameters can be obtained in real time, which will provide accurate information for the advanced dynamics control system and the intelligent chassis integration control. With the introduction of the additional sensing signals and position information, the lane change, overtaking and car-following, and a series of autonomous operations will be achieved \cite{hua2020research}.

Autonomous driving(AD) integrates the information from the surroundings and the vehicle's states via advanced sensing systems (e.g., LIDAR, cameras, GPS, and IMU) \cite{liu2022yolov5, xu2022v2x, xiong2019imu, xu2022cobevt, liu2020vision}, and communication networks \cite{xu2021opencda} to replace or assist the driver. In the last decade, with the development and advances in sensors and computer technologies, AD-related research have become hot topics in both path planning and autonomous tracking control, and have been conducted to ensure vehicle safety, comfort, and energy efficiency \cite{hua2021surrogate, yan2022multi}. 

Many studies have been dedicated to solving the collision-free path planning problem to meet the constraints of collision avoidance, steering speed, and road curvature under continuous conditions \cite{oroko2022obstacle}. State lattice is a graph theory-based approach that has emerged in recent years, where kinematic and dynamical constraints can be regarded through state space rules and repeatable sampling in autonomous driving. Furthermore, unstructured road environments give more possibilities for employing this method. Lattice edges are computed offline to achieve real-time online planning using look-up tables to reduce the computational burden. However, several problems exist for urban areas with highly variable structured environments, such as the discretization of heading angles that may lead to oscillations between two oriented samples \cite{xiong2020imu, bergman2019improved}. A spline-based search tree utilizes a search tree to generate a path that is tangent to all objects and assumes that the optimal path is oriented straight ahead or touches at least one object. Each path segment from one object to the next is defined by a constant acceleration, which is effective in certain scenarios. However, transient lateral acceleration and sudden changes make this approach infeasible for vehicle robust control \cite{rybus2022optimal}. Alternatively, model predictive control (MPC) \cite{ cheng2019longitudinal} has been widely applied in the field of automotive, where the optimization goal is to find the optimal value of a series of curvature changes to minimize the steering effort defined by the road curvature in a collision scenario, along a prediction domain from a fixed sampling time \cite{lu2022real}. A potential field-based path planner is presented to provide obstacle avoidance with an adaptive multi-speed scheduler using a fuzzy system \cite{wahid2020vehicle}. In \cite{wu2019fast}, a time-horizon based MPC method is proposed to reduce the calculation load of vehicle speed predictive control. And the performance is verified through real road simulation experiments.

Nowadays, the vehicle models adopted for autonomous tracking control are divided into three types: geometric, kinematic, and dynamic. The mainstream control algorithms include PID control, sliding mode control (SMC), neural network control, and MPC \cite{snider2009automatic} to obtain the corresponding control parameters: the steering wheel angle, throttle opening, and braking strength \cite{9737534}. There is a certain delay in the actual vehicle control when executing the control command immediately, which is the biggest problem of PID algorithm in autonomous control \cite{cervantes2001pid}; in addition, advanced control algorithms, such as SMC and neural networks, are also widely employed \cite{li2019robust, manzoor2020trajectory}.  However, these algorithms greatly depend on parameter sensitivities and environment information, and different vehicle constraints need to be considered when the vehicle model is embedded; there are great limitations for these optimization algorithms with constraints, although MPC is capable of solving the constraints problem with high computation cost. Therefore, various tracking control algorithms need to be traded off in all aspects. The main challenge with current tracking control algorithms is a reasonable simplification of the vehicle dynamics modeling to improve the real-time performance of the algorithm while ensuring tracking accuracy.

Unfortunately, most algorithms mainly attempt to find collision-free paths with no guarantee of feasibility in the real world due to poor quality and small turns. To mitigate the above limitations in the harsh condition, in this paper, a complicated path with considerable curvatures will be created, and the precise and independent control of four-wheel angles and motor torque can be fully utilized to achieve obstacle-avoiding tracking control based on the multi-degree-of-freedom control platform of the FDWEV \cite{hua2019hierarchical}. The autonomous tracking control of the FDWEV can provide a theoretical and practical basis for developing AD technology. The main contribution of this paper is fourfold. 

1) Artificial potential field (APF) for path planning through comprehensive comparison is optimized to address the potential unreachable and local optimal problems with the improved repulsive force function and the direction of repulsive force. 

2) For the high maneuverability of the FDWEV, the relevant kinematics and dynamic constraints are considered to be integrated into the vehicle speed planning. Then interpolation and piecewise fitting are conducted to provide achievable position information for tracking control.

3) Based on fuzzy control theory, a forward-looking driver behavior model with variable preview distance is designed. The curvature calculation is proposed based on the vehicle lateral dynamics and kinematic models. For the shortcoming of the tracking control algorithm based on curvature calculation, the MPC algorithm is designed by considering the vehicle constraints, real-time prediction, and feedback optimization of the vehicle states.

4) Based on the path planning and vehicle speed planning in the prescribed park, the autonomous tracking controls are verified based on the co-simulation of Simulink/Carsim/Amesim platform under different conditions with speeds of 30 km/h and 50 km/h in a harsh environment with considerable curvatures.

 The rest of this paper is organized as follows: Section 2 formulates the lateral dynamics model. In section 3, path planning with APF method is proposed with the generated reachable points of vehicle tracking control. In Section 4, tracking control based on the curvature method and MPC algorithm are proposed according to the kinematic and dynamics model. Section 5 conducts simulation experiments and analyzes the results for validation and evaluation, followed by the conclusions in Section 6.


\section{Lateral dynamics model}

\begin{figure}[htbp]
\centerline{\includegraphics[width=\columnwidth]{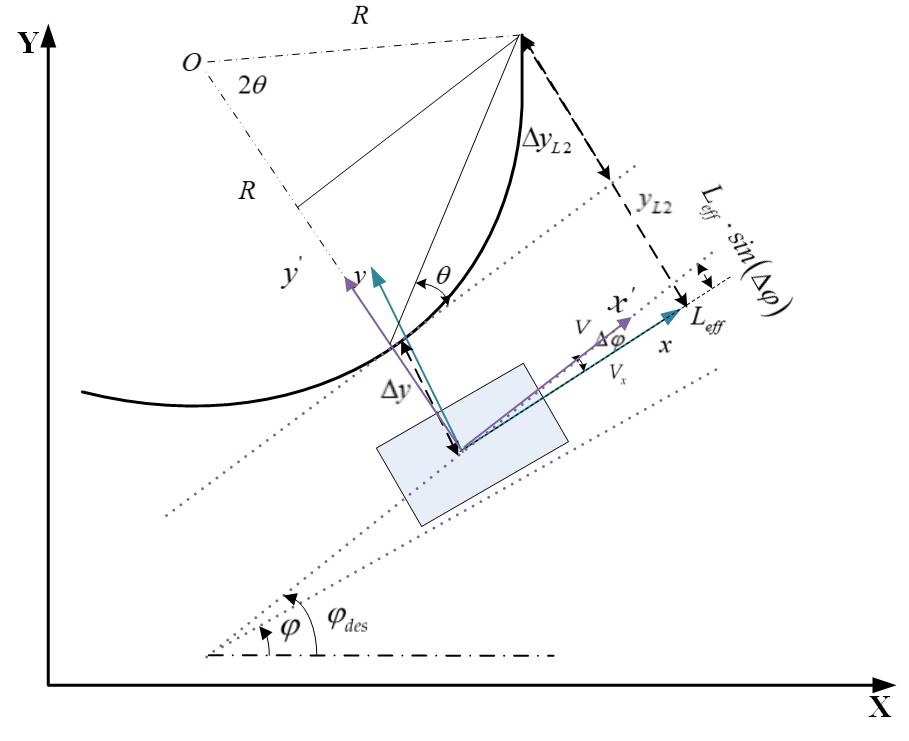}}
\caption{Schematic diagram for lateral dynamics model.}
\label{fig0}
\end{figure}

The lateral dynamics model is established firstly to design the tracking controller. The global coordinate system XY and the vehicle coordinate system xy are defined as shown in Fig. \ref{fig0}. 

In detail,  x is along the vehicle's longitudinal direction, and y is along the lateral direction, a fixed coordinate system $x'y'$ is defined, where $x'$ is defined in the direction of the tangential velocity V and $y'$ is pointing at the rotation center O of the vehicle from the position of the mass center c.g of the vehicle;  the yaw angle of the x-direction of the vehicle coordinate system with respect to the global X-axis direction is denoted by the $\varphi$; V is the tangential velocity of the vehicle, and $V_x$ (called longitudinal velocity) is set as the x-axis component of V; $\delta_f$ is the front wheel steering angle, $\varphi_{des}$ is the desired yaw angle of the vehicle, and $\Delta\varphi$  is the yaw angle deviation with respect to the desired path. In addition, the lateral position error from the vehicle center of mass along the y-direction to the corresponding desired path point is $\Delta y$, $\Delta y$, also known as the current deviation \cite{liu2018intelligent}. Assuming that the desired path has a constant curvature and the vehicle has a constant speed in the longitudinal direction. Based on the two-degree-of-freedom model derived above, the lateral motion control at a constant speed $\Ddot{y} = \Dot{V}_y,\Dot{y} = V_y$, are expressed as follows:
\begin{equation}
\Ddot{y} = -(\frac{C_f + C_r}{V_xm})\Dot{y} + (-V_x - \frac{C_fl_f - C_rl_r}{V_xm})\Dot{\varphi} + \frac{C_f}{m}\delta_f
\label{eq}
\end{equation}
\begin{equation}
\Ddot{\varphi} = -(\frac{C_fl_f - C_rl_r}{I_zV_x})\Dot{y} - (\frac{C_fl_f^2 + C_rl_r^2}{I_zV_x})\Dot{\varphi} + \frac{C_fl_f}{I_z}\delta_f
\label{eq}
\end{equation}
\begin{equation}
\Delta\varphi = \varphi_{des} - \varphi
\label{eq}
\end{equation}
\begin{equation}
\Dot{\Delta}\varphi = \Dot{\varphi}_{des} - \Dot{\varphi}
\label{eq}
\end{equation}
\begin{equation}
\Dot{\Delta}y = -\Dot{y} + V_x\Delta{\varphi}
\label{eq}
\end{equation}

The preview deviation $y_{L2}$ is the lateral position deviation of the vehicle at the effective preview distance $L_{eff}$ from the vehicle center of mass, as shown in Fig. 1. The preview deviation $y_{L2}$ can be considered as the sum of three different distances, which is expressed as follows:
\begin{equation}
y_{L2} = \Delta y_{L2} + \Delta y + L_{eff}\sin(\Delta\varphi)
\label{eqyl2}
\end{equation}
where $\Delta y$ is the current lateral deviation, $\Delta\varphi$ is the heading angle error, and $\varphi$ is the yaw rate. And the only variable $y_{L2}$ depends on the change of the future path. The current lateral position deviation $\Delta y$ depends on the current lateral position of the vehicle, and the third term of Eq. (\ref{eqyl2}) depends on the heading angle of the path and the current heading angle of the vehicle. For small angles $\Delta\varphi$, the above equation can be approximated as:
\begin{equation}
y_{L2} \approx \Delta y_{L2} + \Delta y + L_{eff}\sin(\Delta\varphi)
\label{eq}
\end{equation}

To calculate the preview deviation precisely, it is necessary to find the relationship between the variable $y_{L2}$ and the other variables. If the vehicle travels with a constant tangential velocity $V$ on a circular path with radius $R$, the relationship with the ideal yaw rate $\Dot{\varphi}_{des}$ is expressed as:
\begin{equation}
\Dot{\varphi}_{des} = \frac{V}{R}
\label{eq8}
\end{equation}

It can also be seen from Fig. \ref{fig0} that the relationship between the radius of $R$ and the effective preview distance $L_{eff}$ is:
\begin{equation}
\sin(2\theta) = \frac{L_{eff}}{R}
\label{eq9}
\end{equation}
Where $\theta$ is the angle between the current driving direction of the vehicle and the point $(L_{eff},\Delta y'_{L2})$ defined in the coordinate system $x'y'$. When combining Eq. (\ref{eq8}) and Eq. (\ref{eq9}) and $\theta = \alpha \tan(\frac{\Delta y'_{L2}}{L_{eff}})$ (as can be seen in Fig. \ref{fig0}), the following relationship is derived as follows:
\begin{equation}
\begin{split}
\Dot{\varphi}_{des} &= \frac{V\sin(2\theta)}{L_{eff}}\\
&=\frac{V\sin(2\cdot \alpha\tan(\Delta y'_{L2}/L_{eff}))}{L_{eff}}\\
&\approx \frac{2V\Delta y'_{L2}}{L^2_{eff}}
\label{eq}
\end{split}
\end{equation}

Then 
\begin{equation}
\Delta y'_{L2} \approx \frac{L^2_{eff}\Dot{\varphi}_{des}}{2V}
\label{eq}
\end{equation}
where the distance $\Delta y'_{L2}$ is not equal to $\Delta y_{L2}$, since there is the lateral angle $\alpha$ between the driving direction of the vehicle body and the movement direction of the actual wheels due to the side slip. Therefore, it can be approximated as follows:
\begin{equation}
\Delta y'_{L2} \approx \Delta y_{L2} - L_{eff}\tan(\alpha)
\label{eq}
\end{equation}

As for the small side slip, the lateral angle $\alpha$ can be expressed as:
\begin{equation}
\alpha \approx \tan(\alpha) = \frac{\mathrm{d}y}{\mathrm{d}x} = \frac{\mathrm{d}y}{\mathrm{d}t}\cdot\frac{\mathrm{d}t}{\mathrm{d}x} = \frac{\Dot{y}}{V_x}
\label{eq}
\end{equation}

In addition, the preview deviation $y_{L2}$ needs to be compensated for the side slip. The compensated preview deviation is described as:
\begin{equation}
\begin{split}
\Delta y_{L2Slipcomp} &\approx \Delta y_{L2} - L_{eff}\cdot \frac{\Dot{y}}{V_x}\\ 
&\approx \Delta y_{L2} - \frac{\Dot{y}}{\delta_f V_x}L_{eff}\cdot \delta_f\\
& \approx \Delta y_{L2} - \beta \cdot L_{eff}\cdot\delta_f
\label{eq}
\end{split}
\end{equation}

\begin{equation}
\begin{split}
&\beta = G_{\delta_f,\Dot{y}}(0, V_x) = \frac{\Dot{y}}{\delta_fV_x}\\
&= \frac{C_f V_x(-l_fmV^2_x + C_rL^2_r + C_rl_fl_r)}{-C_fmV^2_xl_f + C_rmV^2_xl_r+C_fC_rl^2_f+2C_fC_rl_fl_r+C_fC_rl^2_r}
\label{eq}
\end{split}
\end{equation}

When the vehicle is in steady state $\Ddot{y} = 0, \Ddot{\varphi} = 0$, the gain of the above equation can be derived based on the two-freedom model and the expression for the preview deviation considering the vehicle side slip can be obtained as:
\begin{equation}
\begin{split}
&y_{L2Slipcomp} = \Delta y_{L2Slipcomp} +\Delta y + L_{eff}\Delta \varphi \\
&=\frac{L^2_{eff}\Dot{\varphi}_{des}}{2V} - \beta\cdot L_{eff}\cdot\delta_f + \Delta y + L_{eff}\Delta\varphi
\label{eq}
\end{split}
\end{equation}


The current lateral deviation $\Delta y$, the lateral velocity $y$, the yaw angle deviation $\Delta \varphi$, and the yaw rate $\varphi$ are set as the state variables $x$; the input variables $u$ are the front wheel angle $\delta_f$ and the desired yaw rate $\varphi_{des}$, and then the output variables are current lateral deviation $\Delta y$, the yaw angle deviation $\Delta \varphi$, and the preview deviation $y_{L2slipcomp}$ considering vehicle side slip. Assuming $V \approx V_x$. Thus, the states equations can be described as follows:
\begin{equation}
\begin{split}
  &\frac{\mathrm{d}}{\mathrm{d}t}\begin{bmatrix} &\Delta y \\ &\Dot{y} \\ &\Delta \varphi \\ &\Dot{\varphi}\end{bmatrix} \\
  &= \begin{bmatrix}
    &0 &-1 &V_x &0\\ &0 &-\frac{C_f + C_r}{V_xm} &0 &-V_x - \frac{C_fl_f - C_rl_r}{V_xm}\\ &0 &0 &0 &-1\\ &0 &-\frac{C_fl_f - C_rl_r}{I_zV_x} &0 &-\frac{C_fl^2_f + C_rl^2_r}{I_zV_x}
\end{bmatrix}\begin{bmatrix}
    &\Delta y\\ &\Dot{y}\\ &\Delta\varphi\\ &\Dot{\varphi}
\end{bmatrix}\\
&+\begin{bmatrix}
    &0 &0\\ &\frac{C_f}{m} &0 \\ &0 &1\\ &\frac{C_fl_f}{I_z} &0
\end{bmatrix}
\begin{bmatrix}
    &\delta_f\\ &\Dot{\varphi}_{des}
\end{bmatrix}  
\end{split}
\label{eq}
\end{equation}

\begin{equation}
\begin{split}
    \begin{bmatrix}
        &\Delta y\\ &\Delta\varphi \\ &y_{L2slipcomp}\\ &\Dot{\varphi}
    \end{bmatrix} &= \begin{bmatrix}
        &1 &0 &0 &0\\ &0 &0 &1 &0\\ &1 &0 &L_{eff} &0 \\ &0 &0 &0 &1
    \end{bmatrix}\begin{bmatrix}
        &\Delta y\\ &\Dot{y} \\ &\Delta\varphi \\ &\Dot{\varphi}
    \end{bmatrix}\\
    &+\begin{bmatrix}
        &0 &0\\ &0 &0\\ &-L_{eff}\beta & \frac{L^2_{eff}}{2V_x}\\ &0 &0
    \end{bmatrix}
    \begin{bmatrix}
        &\delta_f\\ &\Dot{\varphi}_{des}
    \end{bmatrix}
\end{split}
\label{eq}
\end{equation}

The lateral dynamics model is established and used to design the tracking controllers, followed by path planning to provide the tracking input information.

\section{Path planning}
\begin{figure}[htbp]
\centerline{\includegraphics[width=\columnwidth]{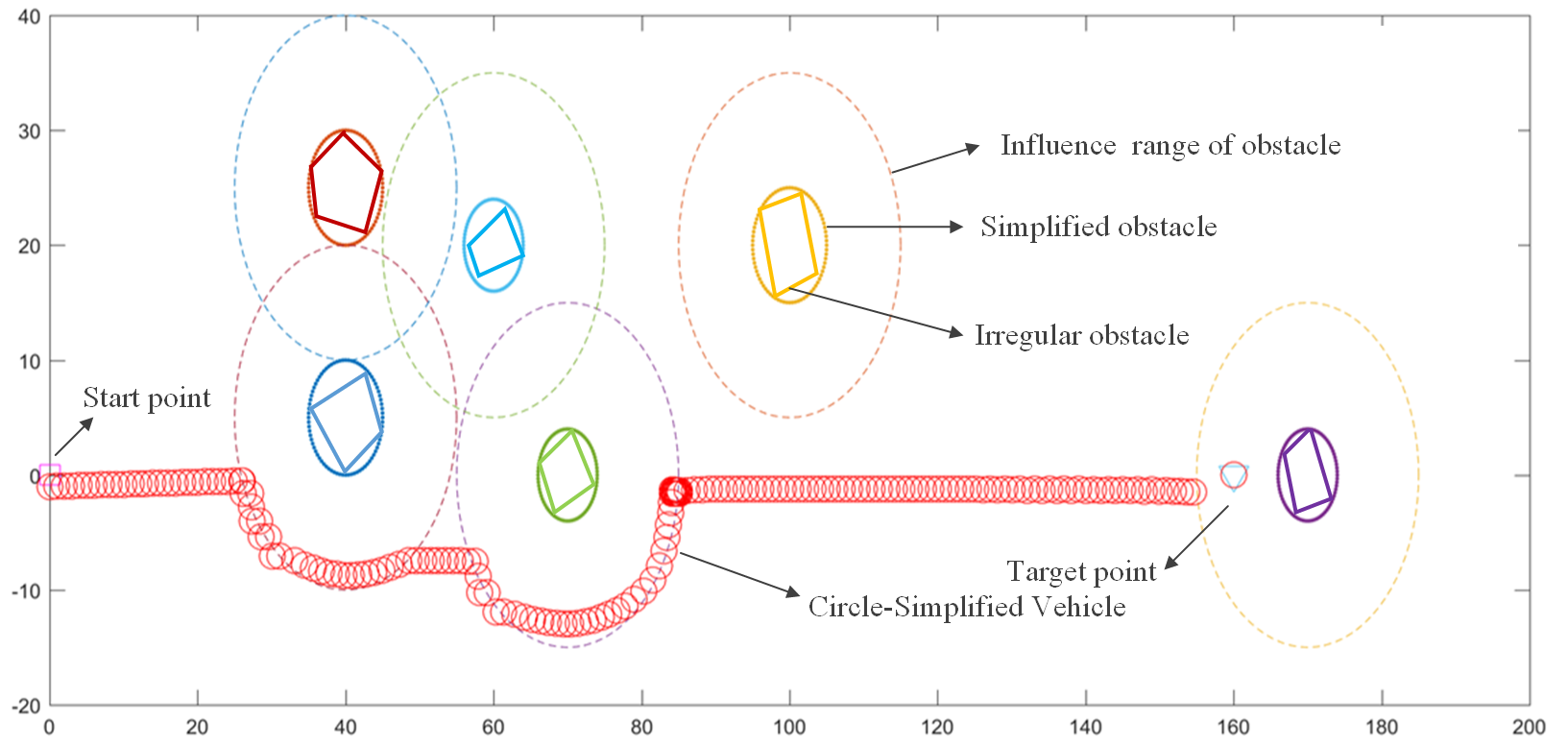}}
\caption{Diagram of the simplified obstacle avoidance in a complicated unstructured scenario}
\label{fig3}
\end{figure}

This section focuses on the path planning of the FDWEV in the presence of static obstacles to provide path point information for tracking control by avoiding obstacles reasonably. Because of the varying sizes and shapes of the various barriers, it is necessary to project the obstacles onto the ground to create a two-dimensional environment. Then, an external circle is utilized to simplify the obstacles so that they fit within a circle. Within the confines of a predetermined park, the vehicle is initially simplified to drive in the form of a circle, with the origin being the mass center of the vehicle. The corresponding diagram is presented in Fig. \ref{fig3}.

\subsection{Path generator}
The artificial potential field (APF) method is widely used for the path generator. However, it currently suffers from the local optimal solution problem and target unreachability. Based on it, some scholars have proposed some improved algorithms\cite{patle2019review, jayaweera2020dynamic}. In this paper, modifying the repulsive force function and repulsive direction have been presented.

(1) Repulsive force function optimization

When the moving vehicle, the target point, and the obstacle are all in the same line and the target point is close to the obstacle, the attractive force of the target point to the moving vehicle decreases while the repulsive force of the obstacle from the moving vehicle increases. As a result, the moving vehicle cannot reach the target point since the repulsive force does not vary as the driving vehicle gets closer to the target point in this scenario. Therefore, modifying the repulsive force function decreases the repulsive force as the moving vehicle gets closer to the target point. In this way, the repulsive and the attractive force are equal to zero at the target point, and the vehicle can maintain the target point.
The improved repulsive field function is described as follows:
\begin{equation}
U_{rep,j}=\left\{
\begin{aligned}
&\frac{1}{2}\eta_j(\frac{1}{d_j(x)} - \frac{1}{Q^*_j})^2 d^2(x,x_{goal}) &d_j(x)\leq Q^*_j\\
&0 &d_j(x) > Q^*_j
\label{eq}
\end{aligned}
\right.
\end{equation}
where $U_{rep,j}$ denotes the improved repulsive force function, $\eta_j$ is the coefficient constant of the repulsive force function. $d_j(x)$ is the shortest distance between the moving vehicle and the $j_{th}$ obstacle. $Q^*_j$ is the maximum range of the influence for $j_{th}$ obstacle on the moving vehicle. When the shortest distance between the moving vehicle and the obstacle is less than  $Q^*_j$, the repulsive force function will affect the moving vehicle. $d(x,x_{goal})$ denotes the shortest distance between the moving vehicle and the target point.

The corresponding repulsive force function is:
\begin{equation}
F_{rep} = -grad(U_{rep,j}) = \left\{
\begin{aligned}
&F_{rep1}N_1 + F_{rep2}N_2 &d_j(x)\leq Q^*_j\\
&0 &d_j(x) > Q^*_j
\label{eq}
\end{aligned}
\right.
\end{equation}
where:
\begin{equation}
F_{rep1} = \eta_j(\frac{1}{\mathrm{d}_j(x)} - \frac{1}{Q^*_j})\frac{\mathrm{d}^2(x,x_{goal})}{\mathrm{d}^2_j(x)}
\label{eq}
\end{equation}
\begin{equation}
F_{rep2} = \eta_j(\frac{1}{\mathrm{d}_j(x)} - \frac{1}{Q^*_j})^2 \mathrm{d}(x,x_{goal})
\label{eq}
\end{equation}
 $F_{rep1}$ and $F_{rep2}$ are the two splitting forces of $F_{rep}$, where the direction of $F_{rep1}$ is from the center of the obstacle pointing to the moving vehicle, the direction of $F_{rep2}$ and the attractive force are the same.
 
This improved method incorporates an adjustment factor $ \mathrm{d}(x,x_{goal})$ so that the repulsive force decreases while the moving vehicle is approaching the target point. The repulsive force is zero, and the attractive force is also zero when reaching the target point.

(2) Repulsive force direction reformation

Altering the repulsive force described above is insufficient to solve either of the difficulties. As a result, based on the repulsive force function, modifying the direction of repulsive force $F_{rep1}$ has been conducted. $F_{rep1}$ and $F_{rep2}$ are the two splitting forces of the repulsive force, and the direction of $F_{rep1}$ is from the obstacle to the moving vehicle, and the direction of $F_{rep2}$ is from the moving vehicle to the target point. When the angle between $F_{rep2}$ and the attractive force is larger than $90^{\circ}$, this may nevertheless result in a locally optimal solution. Therefore, modifying the direction of the repulsive force is to define $F_{rep1}$ as a tangent direction along the influencing range of the obstacle and the angle between it and the attractive force, which is less than or equal to $90^{\circ}$. 

\subsection{Speed planning}

(1) Vehicle dynamics constraints

According to the vehicle dynamics theory analysis, the under-steering gain can be defined as $K=\frac{m}{(l_f+l_r)^2}(\frac{l_f}{C_{\alpha r}} - \frac{l_r}{C_{\alpha f}})$. Then, steering gain is the ratio of the yaw rate to the front wheel angle, which is $\frac{\varphi}{\delta_f} = \frac{V_x/(l_f+l_r)}{1+KV^2_x}$, ensuring the vehicle's stability and safety.\cite{liu2017vehicle}.
\begin{equation}
\left\{
\begin{aligned}
&\frac{\Dot{\varphi}}{\delta} = \frac{V_x}{l_f + l_r} \\
& V_x = \Dot{\varphi}R
\label{eq}
\end{aligned}
\right.
\end{equation}

In the vehicle coordinate system, the vehicle lateral acceleration can be expressed as:

\begin{equation}
\alpha_y = \lim_{\Delta t \to 0}\frac{\Delta V_y + V_x\cdot \Delta\theta}{\Delta t}
\label{eq}
\end{equation}

To avoid obstacles, the vehicle is steering in such a way as to satisfy steady-state steering and to make certain that the lateral side slip angle of the center of mass is as small as possible. Assuming the velocity of the vehicle along the longitudinal axis remains constant, the lateral acceleration can be represented as follows:
\begin{equation}
\alpha_y = \frac{V^2_x}{l_f+l_r}\cdot \delta
\label{eq}
\end{equation}

To make the tires in the linear area, the lateral acceleration should be no more than 0.4 g, and the front wheel needs to be satisfied as follows:
\begin{equation}
\delta \leq 0.4g\cdot\frac{l_f+l_r}{V^2_x}
\label{eq}
\end{equation}

Therefore, the vehicle's steering would be simultaneously impacted by the velocity and steering radius. And the front wheel steering angle should meet smoothly and securely:
\begin{equation}
\delta = \min(0.4g\cdot\frac{l_f+l_r}{V^2_x}, \frac{l_f+l_r}{R_{\min}})
\label{eq}
\end{equation}

During steering, the longitudinal speed is far too high to have any impact on the safety of the vehicle, and the vehicle speed at any given curve can be satisfied:
\begin{equation}
V_{\lim} = \sqrt{g\mu/\rho_d}
\label{eqlimit}
\end{equation}

Eq. \ref{eqlimit} the limited speed under ideal conditions, it is required to adjust the vehicle speed by inserting an adjustment factor $\lambda_d$ by considering the real instantaneous scenario:
\begin{equation}
V_{\lim} = \lambda_d\sqrt{g\mu/\rho_d}
\label{eq}
\end{equation}

It can be seen that the only factors that influence the limited speed are the radius of the road, denoted by $\lambda_d$, and the maximum adhesion coefficient, denoted by $\mu$. Then the longitudinal reference speed should also consider the maximum legal speed restriction. Therefore, the following can be obtained\cite{gao2019multi} :
\begin{equation}
V_{ref} = \min(V_{set}, V_{\lim})
\label{eq}
\end{equation}
where $V_{set}$ is the desired target speed.

(2) Vehicle kinematic constraints

The front wheels need to be in accordance with the ideal Ackermann geometry steering, and the vehicle kinematic model can be described as follows.
\begin{equation}
\left\{
\begin{aligned}
&\Dot{X} = V\sin\theta \\
&\Dot{Y} = V\cos\theta \\
&\theta = \frac{V\tan\delta_f}{l_f+l_r}
\label{eq}
\end{aligned}
\right.
\end{equation}

The incomplete constraint can be expressed as:
\begin{equation}
\left\{
\begin{aligned}
&\Dot{X} = \Dot{X}\sin\theta - \Dot{Y}\cos\theta \\
&\Dot{Y} = \Dot{X}\sin(\delta_f + \theta) - \Dot{Y}\cos(\delta_f + \theta) - (l_f+l_r)\theta\cos\delta_f
\label{eq}
\end{aligned}
\right.
\end{equation}
where $\theta$ is the angle between the vehicle and the X-axis direction. Based on the above constraints, it is clear that the kinematic constraints include a speed limit $V_{\lim}$ without side slip, with different curvatures:
\begin{equation}
\begin{split}
&V_{\lim} = \sqrt{\mu g\sqrt{(1 + (l_f+l_r)^2\rho^2_d)}\cdot\sqrt{R^2 - (l_f+l_r)^2}}\\
&\rho_d = \frac{1}{R}
\label{eq}
\end{split}
\end{equation}

The reference speed $V_{ref}$ will be selected as the smallest speed value.

\section{Tracking control}
The tracking control algorithms are designed to accurately follow the path and the corresponding vehicle speed to avoid static obstacles. Firstly, the fuzzy control is used to obtain the preview time, combined with speed planning to get the variable preview distance. Furthermore, the vehicle kinematics and dynamics models are considered respectively with vehicle constraints. Thus, a tracking control model based on curvature calculation (CC-based) and model prediction control (MPC-based) methods are established.

\subsection{Design of tracking controller based on curvature calculation}
In the lateral dynamics model, the defined preview deviation $y_{L2}$ contains three parts, $\Delta y_{L2}$ indicating the future path change, $\Delta y$ representing the current lateral position deviation, and  $\Delta \varphi$ indicating the current heading angle deviation.

(1) Curvature calculation based on current lateral deviation $\Delta y$

 To calculate and correct the deviation between the current path and the reference path, PID controllers, as the most common feedback controllers used in industrial fields, were first employed due to no representation of the internal system and the straightforward application to determine the ideal curvature $\kappa_{\Delta y}$ based on the present deviance. The following is an equation representing the PID controller.
\begin{equation}
\kappa_{\Delta y}(k) = K_p\Delta y(k) + K_i\sum_{j=0}^k\Delta y(j) + K_d[\Delta y(k) - \Delta y(k-1)]
\label{eq}
\end{equation}

\begin{figure}[htbp]
\centerline{\includegraphics[width=\columnwidth]{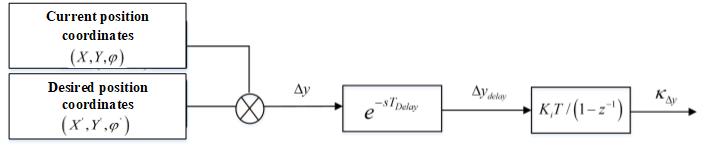}}
\caption{Block diagram of the current deviation control loop}
\label{fig4}
\end{figure}

The $ K_p$ is the proportional coefficient, the $ K_i$ is the integral coefficient, and the $ K_d$ is the differential coefficient. The I controller provides the most effective results by modifying the three parameters in the repeated simulations. As a result, the I controller is used to determine the ideal curvature. Then, the sensor measurement has a certain delay in the real-world case, which results in a delay in the acquired current lateral deviation $\Delta y$. So a delayed module $e^{sT_{Delay}}$ needs to be introduced to provide a more accurate control in Fig. \ref{fig4}.

(2) Curvature calculation based on preview deviation $y_{L2}$

For the control loop of the $y_{L2}$, the data delay is first considered to get the delayed preview deviation $y_{L2delayed}$, and then the derivation from the lateral dynamics model is given:
\begin{equation}
\begin{split}
&\Dot{\varphi}_{des} =  \frac{V\sin \left ( 2\theta  \right )}{L_eff}  \approx \frac{2V{\Delta y{}'}_{L2}}{L^{2}_eff}
\label{eq}
\end{split}
\end{equation}

The preview deviation distance can therefore be converted into the ideal curvature as shown below:
\begin{equation}
\begin{split}
&\kappa_{y_{L2}} =  \frac{\Dot{\varphi}_{des}}{V_x}  \approx \frac{2{\Delta y_{L2}}}{L^{2}_eff}
\label{eq}
\end{split}
\end{equation}

The conversion of the preview deviation into the corresponding curvature value by gain $2/{L^{2}_eff}$ is conducted. Then a parameter $G_{out}$ can be added between the two distances, the range of which is set to $0< G_{out}< 1$ to make the steering less aggressive. Since the vehicle is subject to side slip, a compensation amount is considered to calculate the preview distance deviation to obtain $y_{L2_{slipcomp}}$, which is:
\begin{equation}
\begin{split}
&y_{L2_{slipcomp}}=y_{L2} - \frac{dy}{dx}{L_eff}
\label{eq}
\end{split}
\end{equation}
where
\begin{equation}
\begin{split}
&\frac{dy}{dx}=\frac{dy}{dt}\frac{dt}{dx}=\frac{\dot{y}}{\delta _{wheel}}\frac{1}{V_x}\delta _{f}=\frac{G_{\delta _{wheel},\dot{y}}(0,V_x)}{V_x} \frac{\delta_{steeringwheel}}{SteerRatio} 
\label{eq}
\end{split}
\end{equation}

\begin{figure}[htbp]
\centerline{\includegraphics[width=\columnwidth]{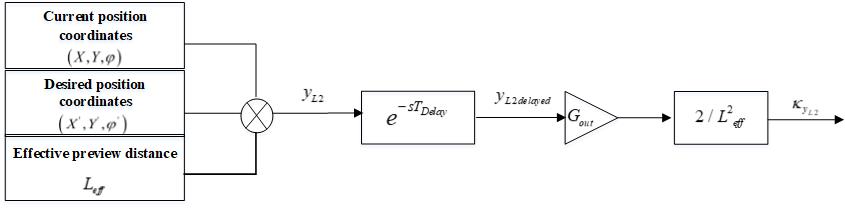}}
\caption{Block diagram of the preview deviation control loop}
\label{fig5}
\end{figure}

The compensated preview distance deviation is brought into the curvature calculation module as the actual preview deviation, which is $y_{L2_{slipcomp}}=y_{L2}$. The control block diagram can be obtained as shown in Fig. \ref{fig5}.

Finally, the ideal curvature $\kappa_{\Delta y}$ based on the current deviation and the ideal curvature $\kappa_{y_{L2}}$ based on the preview distance are transformed into the corresponding steering angle in different cases, the conversion between the steering angle and the ideal curvature is derived as follows:
\begin{equation}
\begin{split}
&G_{curvature,\delta_f}(0,V_x) = \frac{\delta_f}{\kappa} = \frac{\delta_f V_x}{\Dot{\varphi}} = G^{-1}_{\delta_{f,\Dot{\Psi}}}(0,V_x)\cdot V_x\\
&=\frac{-C_fmV^2_xl_f + C_rmV^2_xl_r + C_fC_rl^2_f + 2C_fC_rl_fl_r + C_fC_rl^2_r}{C_fC_rV_x(l_f + l_r)}V_x\\
&=\frac{-C_fmV^2_xl_f + C_rmV^2_xl_r + C_fC_rl^2_f + 2C_fC_rl_fl_r + C_fC_rl^2_r}{C_fC_r(l_f + l_r)}
\label{eq}
\end{split}
\end{equation}

\begin{figure*}[htbp]
\centerline{\includegraphics[width=16cm]{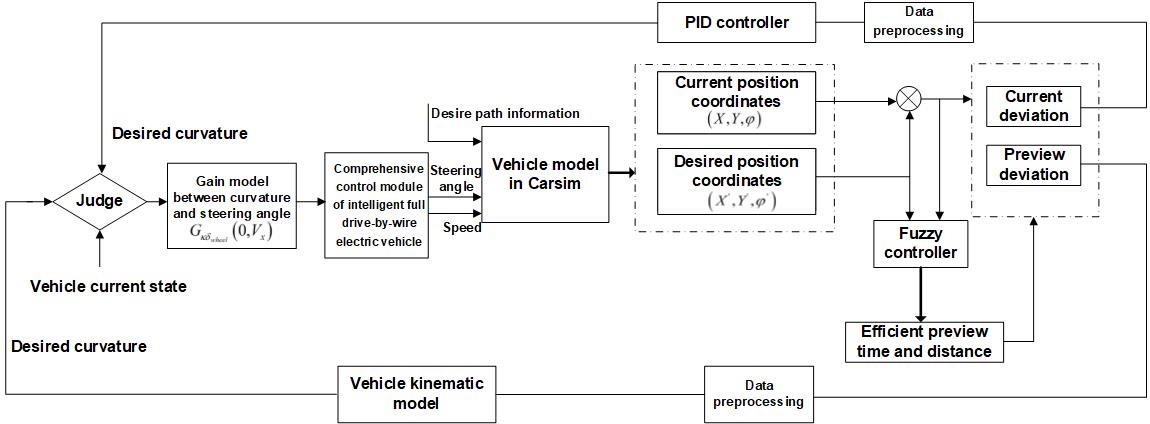}}
\caption{Block diagram of CC-based control strategy design}
\label{fig6}
\end{figure*}

The effective preview time can be obtained from the fuzzy controller and the speed limit obtained from the vehicle dynamics and kinematics. An effective preview distance can be obtained by multiplying the speed with the effective preview time. The CC-based tracking controller framework can therefore be obtained as shown in Fig. \ref{fig6}. $G_{out}$ is an adaptive parameter, and $LocalOffsetLimit$ is set to 0.2m, and the logic pseudo-code for the judgment is:
\begin{algorithm}
    \caption{The logic pseudo-code for the judgment}
    \begin{algorithmic}
        \IF{$ALatLimit \leq 0.4g$}
        \STATE $G_{out} = 0.8$
        \ELSE
        \STATE $G_{out} = 0.65$
        \ENDIF
        
        \IF{$\left|\Delta y\right| > LocalOffsetLimit$}
        \STATE $DisableLocalOffsetIntegrator = True$
        \ELSE
        \STATE $DisableLocalOffsetIntegrator = False$
        \ENDIF
    \end{algorithmic}
\end{algorithm}

\subsection{Design of tracking controller based on model predictive control}
(1) Optimization objectives

The aim is not only to require the vehicle to produce the smallest lateral deviation but also to achieve the smallest steering wheel angle variation based on the consideration of the stability and safety \cite{zhang2017nonlinear, chen2020multiobjective}. Therefore, the optimization objective function is defined as follows:
\begin{equation}
\begin{split}
V(k) &= \sum_{i=1}^{H_p}\left\|z(k + i|k) - r(k+i) \right\|^2_{Q(i)} \\
&+ \sum_{i=1}^{H_p}\left\|u(k+i|k) \right\|^2_{R_u(i)} +\sum_{i=1}^{H_p}\left\|\Delta u(k+i|k)\right\|^2_{Q(i)}
\label{eq}
\end{split}
\end{equation}
The reference path information is defined as:
\begin{equation}
R(k) = \left[ r(k + 1|k)\cdots r(k + H_p|k)\right]^T
\label{eq}
\end{equation}

Normalization is required due to different magnitudes of three terms for the objective function; then the objective function can be expressed as:
\begin{equation}
V(k) = \left\|Z(k) - R(k)\right\|^2_Q + \left\|U(k)\right\|^2_{R_u} + \left\|\Delta U(k)\right\|^2_R
\label{eq}
\end{equation}


The vehicle lateral tracking deviation is defined as:
\begin{equation}
\varepsilon(k) = R(k) - Hx(k) - Pu(k-1)
\label{eq}    
\end{equation}

Bring the lateral tracking deviation into the objective function as follows:
\begin{equation}
\begin{split}
V(k) &= \left\|S\Delta U(k) - \varepsilon(k)\right\|^2_Q + \left\|U(k)\right\|^2_{R_u} + \left\|\Delta U(k)\right\|^2_R\\
&=\left[\Delta U(k)^T S^T - \varepsilon(k)^T\right]Q\left[S\Delta U(k) - \varepsilon(k)\right]\\
&+\left[u(k-1)^T\Gamma^T + \Delta u(k)^T\Lambda^T\right]R_u\left[\Gamma u(k-1) + \Lambda\Delta u(k)\right]\\
&+\Delta U(k)^TR\Delta U(K)
\label{eq}   
\end{split} 
\end{equation}
The above equation is simplified as follows:
\begin{equation}
\begin{split}
&V(k) = C - \Delta U(k)^T M + \Delta U(k)^T N\Delta U(K)\\
&M = 2S^T Q\varepsilon(k) - 2\Lambda^T R_u\Gamma u(k-1)\\
&N = \Lambda^T R_u \Lambda + R + S^T QS
\label{eq}   
\end{split} 
\end{equation}

It is easy to see that M and N are irrelevant to $\Delta U(k)$. Then the dimensions of the input, output, and state variables in the system are assumed to be l, m, and n; therefore, the dimensions of the matrix can be obtained as shown in Table 1.
\begin{table}[h]
\centering
\caption{Table of Matrix dimension}
\label{table}
\begin{tabular}{|c|c|}
\hline
Matrix& 
Dimension \\
\hline
$\Gamma$& $lH_u \times l$\\
$\Lambda$& $lH_u \times lH_u$ \\
$Q$& $mH_p \times mH_p$\\
$R_u$& $lH_u \times lH_u$\\
$R$& $lH_u \times lH_u$\\
$H$& $mH_p \times n$\\
$P$& $mH_p \times l$\\
$S$& $mH_p \times lH_u$\\
$\varepsilon$& $m \times 1$\\
$M$& $lH_u \times 1$\\
$N$& $lH_u \times lH_u$\\
\hline
\end{tabular}
\label{tab1}
\end{table}

(2) Constraints

To ensure the safety and stability of the vehicle, limitations on the steering wheel angle, steering wheel angle rate, lateral acceleration, lateral deviation, and actuator capacity should be considered when designing the control algorithms. Firstly, the constraint of the front wheel angle as the model input is set as $-25^\circ - 25^\circ$, which is the extreme value of the steering angle. Thus, the constraints on the front wheel angle and the corresponding variation are designed as follows:
\begin{equation}
-25^\circ \leq \delta_f \leq 25^\circ\\
\label{eq}   
\end{equation}
\begin{equation}
-0.47^\circ \leq \Delta\delta_f \leq 0.47^\circ
\label{eq}   
\end{equation}

Then the lateral acceleration is $\alpha_y \leq \mu g$ in the theoretical case, and this paper sets the limit as $\alpha_y \leq 0.85\mu g$ for safety. The yaw rate can also be obtained as follows:
\begin{equation}
\begin{split}
&\psi\cdot V_x \leq 0.85 \mu g\\
&\Dot{\psi} \leq \frac{0.85 \mu g}{V_x}
\label{eq}   
\end{split} 
\end{equation}

Finally, the side slip angle cannot be exceeded to $5^\circ$, within which there is a linear relationship between the lateral force and side slip angle according to the tyre characteristics. Therefore, the constraint of the side slip angle of the front wheel is:
\begin{equation}
-2.5^\circ \leq \alpha \leq 2.5^\circ
\label{eq}   
\end{equation}

(3) Optimization with constraints 

The above constraints can first be expressed as linear inequalities as:
\begin{equation}
E\begin{bmatrix}
    &\Delta U(k)\\
    &1
\end{bmatrix} \leq 0
\label{eq}   
\end{equation}
\begin{equation}
F\begin{bmatrix}
    &U(k)\\
    &1
\end{bmatrix} \leq 0
\label{eq53}   
\end{equation}
\begin{equation}
G\begin{bmatrix}
    &Z(k)\\
    &1
\end{bmatrix} \leq 0
\label{eq54}   
\end{equation}
where the matrix $F$ is denoted as $f = [F_1,F_2,\cdots,F_{H_u},f]$, $F_i$ is the matrix of $q \times m$, $f$ is the matrix of $q \times 1$, it can be described as:
\begin{equation}
\sum_{i=1}^{H_u}F_iu(k+i-1|k) + f \leq 0
\label{eq55}   
\end{equation}
Because
\begin{equation}
u(k+i-1|k) = u(k-1) + \sum_{j=1}^{i-1}\Delta u(k+j|k)
\label{eq}   
\end{equation}
Bring this equation into Eq. (\ref{eq55}):
\begin{equation}
\begin{split}
&\sum_{j=1}^{H_u}F_j\Delta u(k|k) + \sum_{j=2}^{H_u}F_j\Delta u(k+2|k)+\cdots\\
&+F_{H_u}\Delta u(k+H_u - 1|k) + \sum_{j=1}^{H_u}F_ju(k-1) + f \leq 0
\label{eq}   
\end{split} 
\end{equation}
where $F_i = \sum_{j=1}^{H_u}F_j$
\begin{equation}
F\Delta U(k) \leq -F_1u(k-1) - f
\label{eq}   
\end{equation}

Thus Eq. (\ref{eq53}) can be transformed into a linear constraint on $\Delta U(k)$. Similarly, Eq. (\ref{eq54}) can be transformed into a constraint on $\Delta U(k)$.
\begin{equation}
G\begin{bmatrix}
    &Hx(k|k) + Pu(k-1) + S\Delta U(K)\\
    &1
\end{bmatrix} \leq 0
\label{eq}   
\end{equation}
where $G=[T,g]$, $g$ is the last column of $G$, since it is obtained:
\begin{equation}
T\left[Hx(k|k) + Pu(k-1)\right] + TS\Delta U(k) + g \leq 0
\label{eq}   
\end{equation}

Furthermore, it can be described as:
\begin{equation}
TS\Delta U(k) \leq -T\left[Hx(k|k) + Pu(k-1)\right] - g
\label{eq}   
\end{equation}

Finally, it can be summarized as:
\begin{equation}
W\Delta U(k) \leq w
\label{eq}   
\end{equation}

Combining with three equations, it can be described as:
\begin{equation}
\begin{bmatrix}
    &E\\ &TS\\ &W
\end{bmatrix}\Delta U(k) \leq
\begin{bmatrix}
    &-E_1u(k-1)-e\\ 
    &-T\left[Hx(k|k) + Pu(k-1)\right] -g\\ 
    &w
\end{bmatrix}
\label{eq}   
\end{equation}

Thus, the optimization problem can be expressed as follows:
\begin{equation}
\begin{split}
  min \qquad &-\Delta U(k)^TM + \Delta U(k)^TN\Delta U(k)\\
  &sub \qquad \lambda\Delta U(k) \leq \kappa
\label{eq}      
\end{split}
\end{equation}

The optimization solution is obtained by solving the gradient function for the objective function and making it equal to 0. 
\begin{equation}
\nabla_{\Delta U(k)}V = -M + 2N\Delta U(k) = 0
\label{eq}    
\end{equation}

So the optimal solution is obtained as follows:
\begin{equation}
\Delta U(k)_{opt} = \frac{1}{2}N^{-1}M
\label{eq}    
\end{equation}

\section{Simulation and Results}
The Simulink/Carsim/Amesim co-simulation is developed under the vehicle speed of 30 km/h and 50 km/h, respectively. Based on the path planning of the tracking control module in the static environment, the target steering angle and vehicle speed are obtained, which will be utilized as the inputs for full drive-by-wire electric vehicles chassis control\cite{chen2019comprehensive}. Following the output of the hierarchical chassis control, the ideal torque of four wheels and four angles will be produced, which are utilized as the inputs of Carsim. This paper will verify the path-tracking control strategies under various working situations.




\subsection{Results of planning and tracking control algorithm based on curvature calculation}
A practicable path within a specific range is designed for the FDWEV, with various obstacles, the start point, and a target point to offer correct position information for the tracking control module. After obtaining the path point information, the generated points will be interpolated with a three-order spline. Then the points are processed using a five-order polynomial fitting for every 10 path points to make them meet the vehicle driving requirements and generate a reasonable path\cite{elhoseny2018bezier}. Then simulation settings with vehicle speeds of 30 km/h and 50 km/h are created in a specific scenario with considerable curvatures to validate the effectiveness of the tracking control algorithm.

(1) Vehicle speed is 30 km/h 

\begin{figure*}
    \centering
    \subfloat[ ]{
    \includegraphics[width=0.32\linewidth]{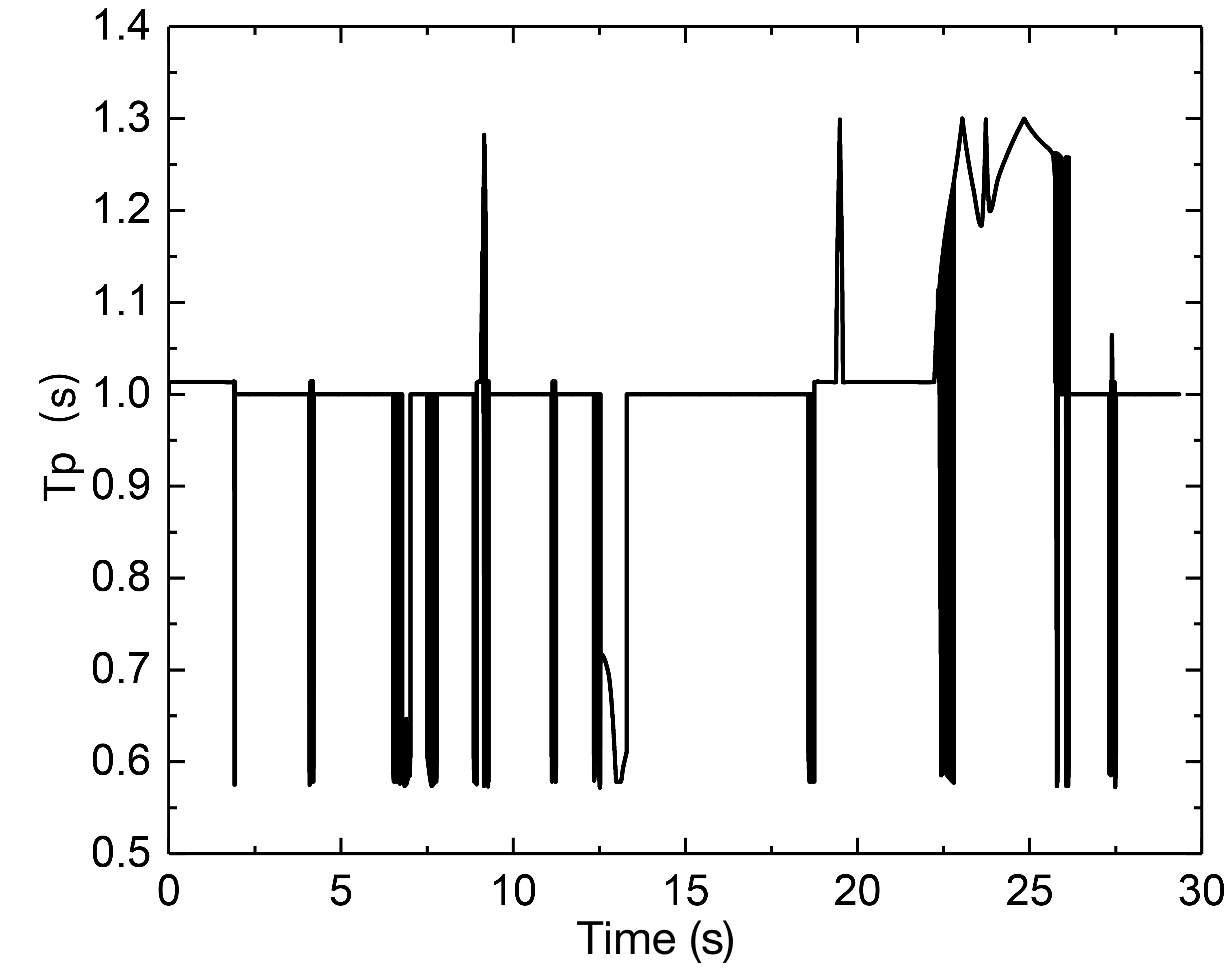}}
    \label{1a}\hfill
    \subfloat[ ]{
    \includegraphics[width=0.3\linewidth]{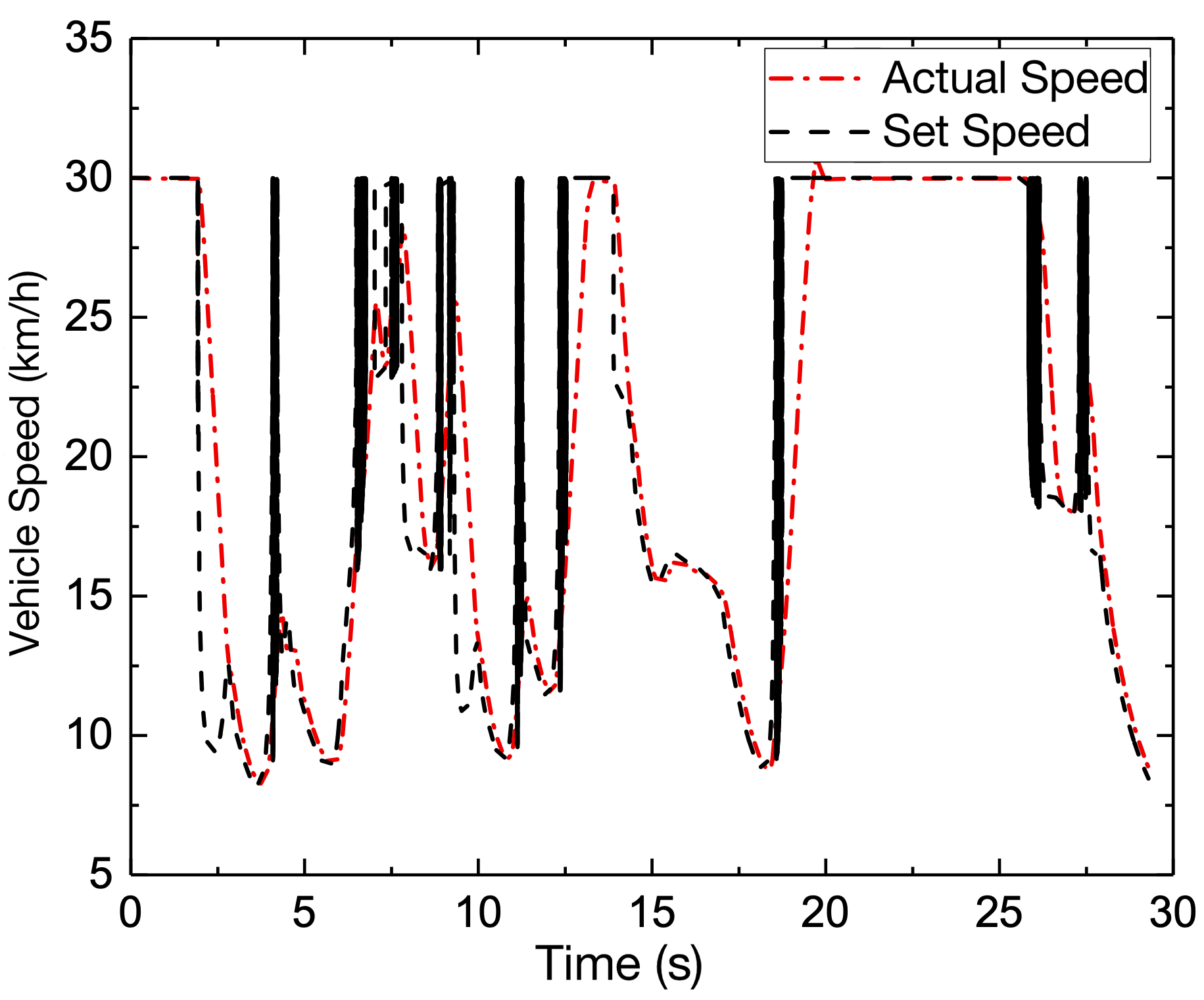}}
    \label{1b}\hfill
    \subfloat[ ]{
    \includegraphics[width=0.33\linewidth]{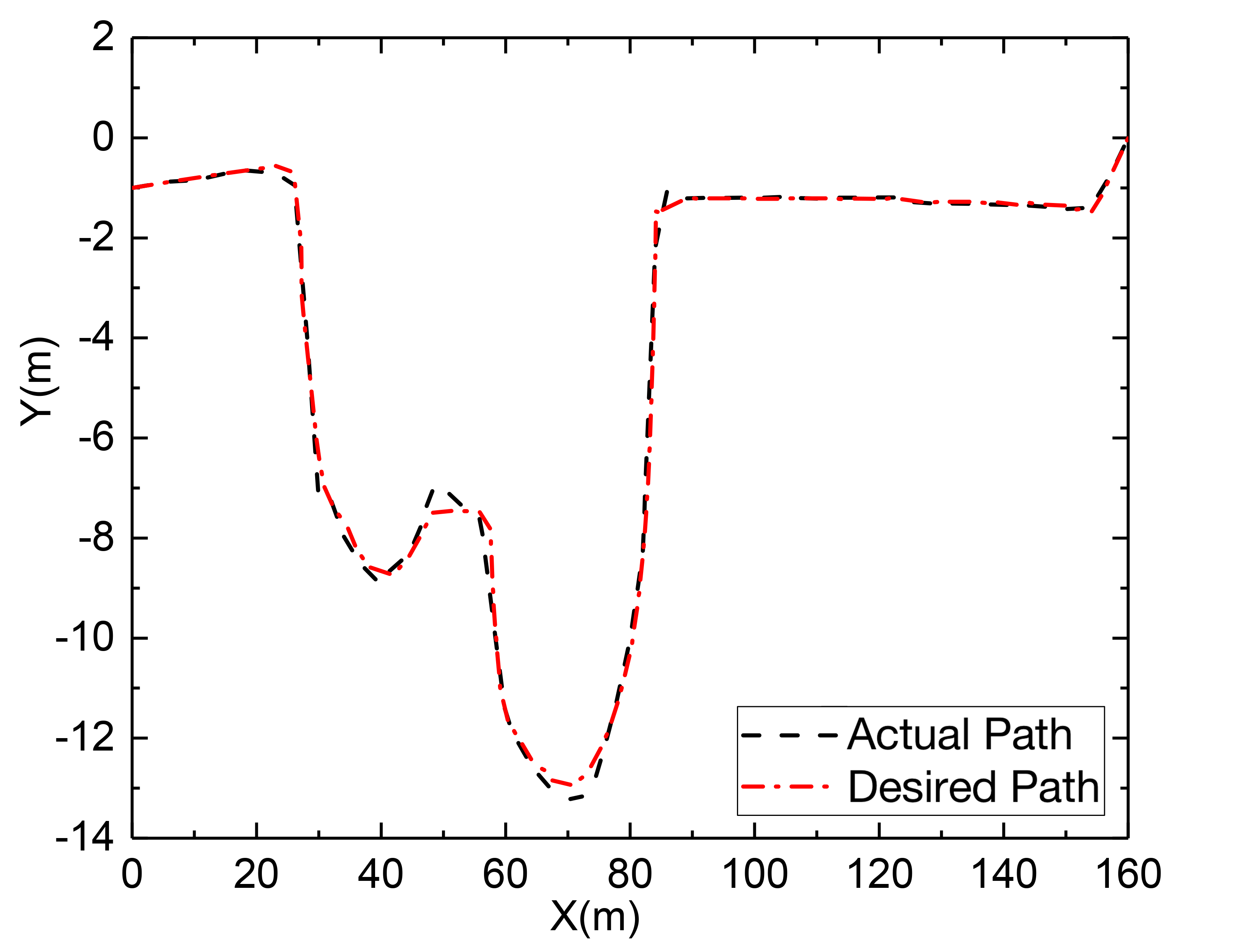}}
    \label{1c}\hfill
    \subfloat[ ]{
    \includegraphics[width=0.33\linewidth]{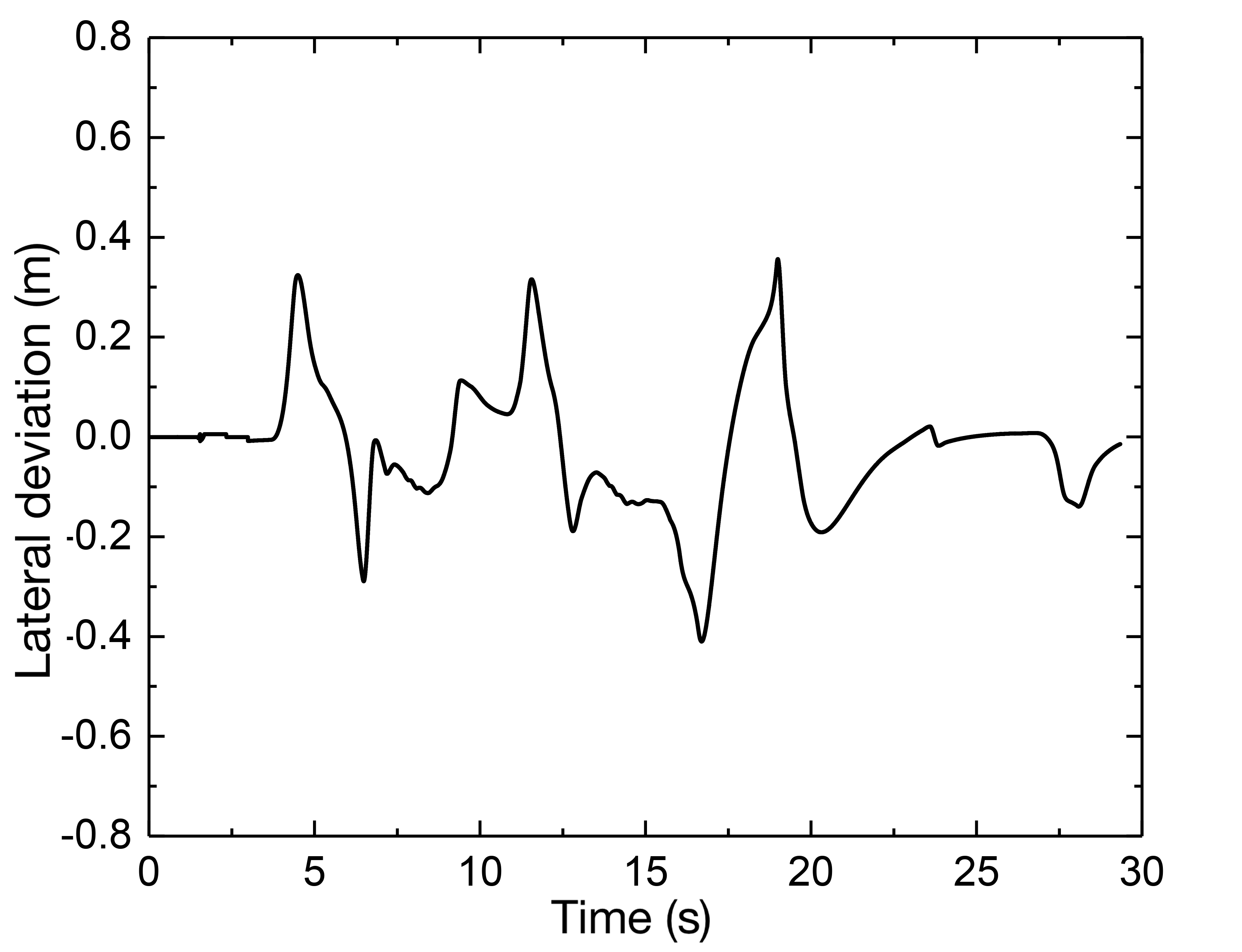}}
    \label{1d}\hfill
    \subfloat[ ]{
    \includegraphics[width=0.33\linewidth]{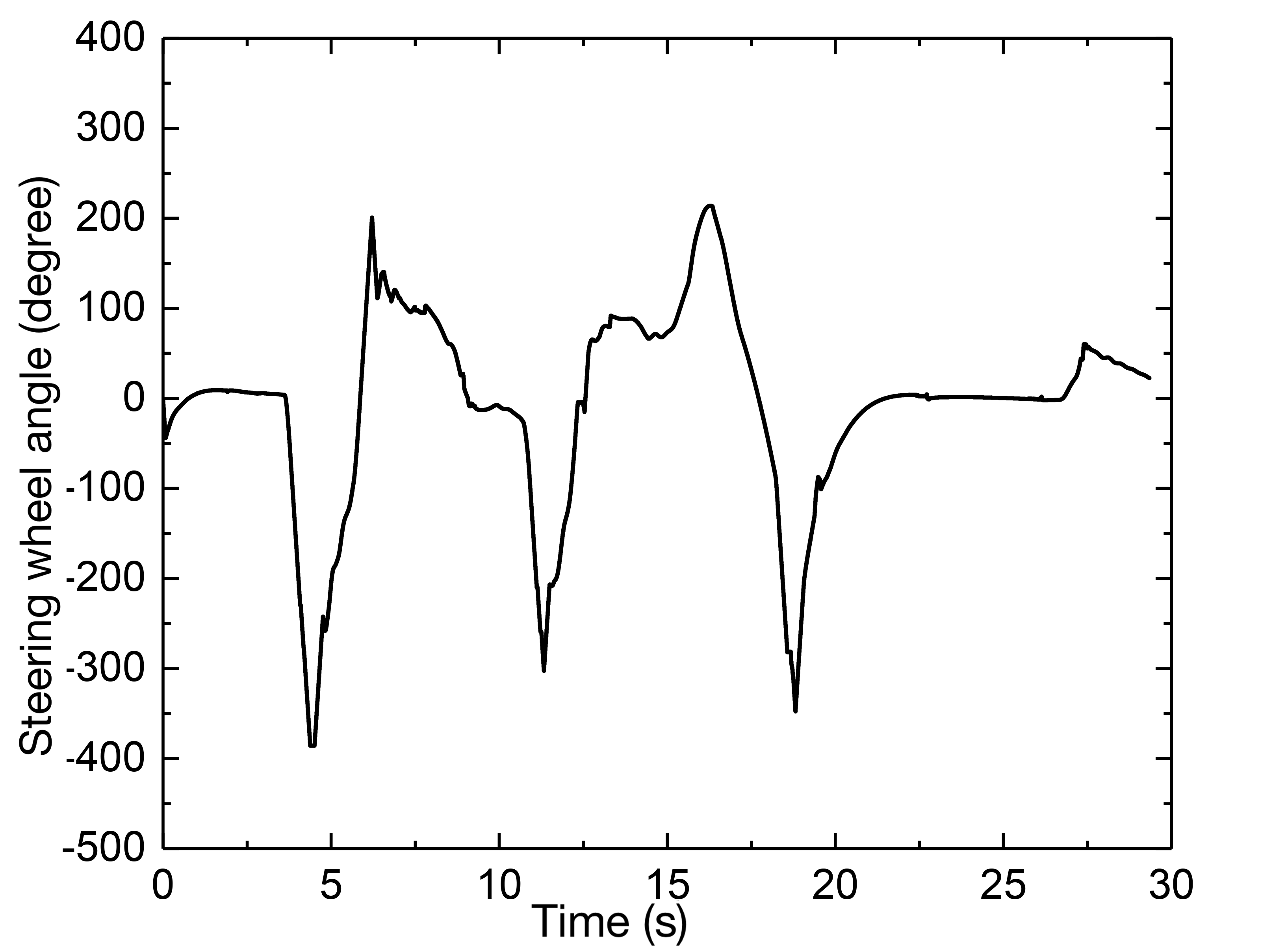}}
    \label{1e}\hfill
    \subfloat[ ]{
    \includegraphics[width=0.3\linewidth]{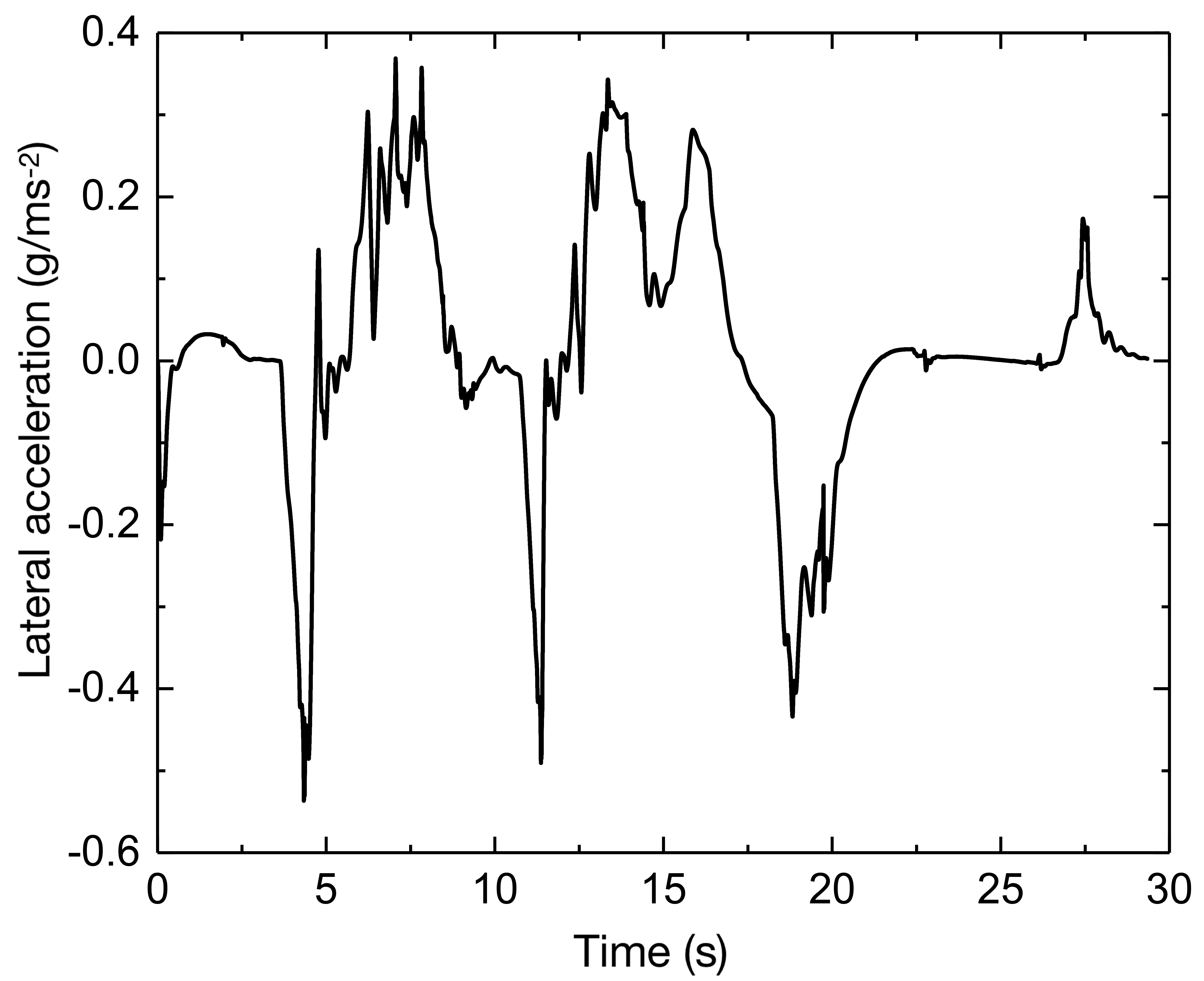}}
    \label{1f}\hfill
    \caption{(a) $T_{p}$ change curve; (b) Speed tracking results; (c) Path tracking results; (d) Lateral deviation results; (e) Corresponding steering wheel angle results; (f) Lateral acceleration results. The setting with CC-based method under 30 km/h.}
    \label{fig7}
\end{figure*}

According to the various deviations and curvature inputs, the output corresponding to the preview time $T_{p}$ can be generated using the fuzzy controller, as illustrated in Fig. \ref{fig7} (a). Based on the speed limitations for the curvature of the road, in this paper, we use the safety factor, which is defined as the adjustment factor $\lambda_d$ = 0.65, and the adhesion coefficient $\mu = 0.85$. As a result, the reference speed curve with the restriction can be obtained, and the speed tracking results show good performance by the chassis integrated control strategy, as shown in Fig. \ref{fig7} (b). 

Fig. \ref{fig7} (c) depicts the path tracking results at a speed of 30 km/h working condition, and Fig. \ref{fig7} (d) depicts the corresponding lateral deviation. To determine the present curvature, a delay is introduced with $T_{Delay} = 0.065s$, and the PID controller computes a curvature value. By repeated simulations, $K_{i} = 1.5$ is eventually determined, implying that the I controller is employed to regulate the deviation directly. A delay with $T_{Delay} = 0.065s$ is also added to determine the preview deviation, the preview time determined by the fuzzy controller can then be multiplied by the reference speed to obtain the variable preview distance and the gain is set as $G_{out}$ = 0.8, which would produce a curvature value with $LocalOffsetLimit$ = 0.2m. As a result, as illustrated in Fig. \ref{fig7} (e), the steering wheel angle can be calculated, where the transmission ratio is chosen as 16. Further, as illustrated in Fig. \ref{fig7} (f) with the lateral acceleration, it shows that the vehicle is mostly in the stable area and the steering curvature is significant, and near to 0.1, the lateral acceleration reaches its maximum, increasing the lateral force and allowing it to negotiate the curve successfully.

(2) Vehicle speed is 50 km/h

\begin{figure*}
    \centering
    \subfloat[ ]{
    \includegraphics[width=0.32\linewidth]{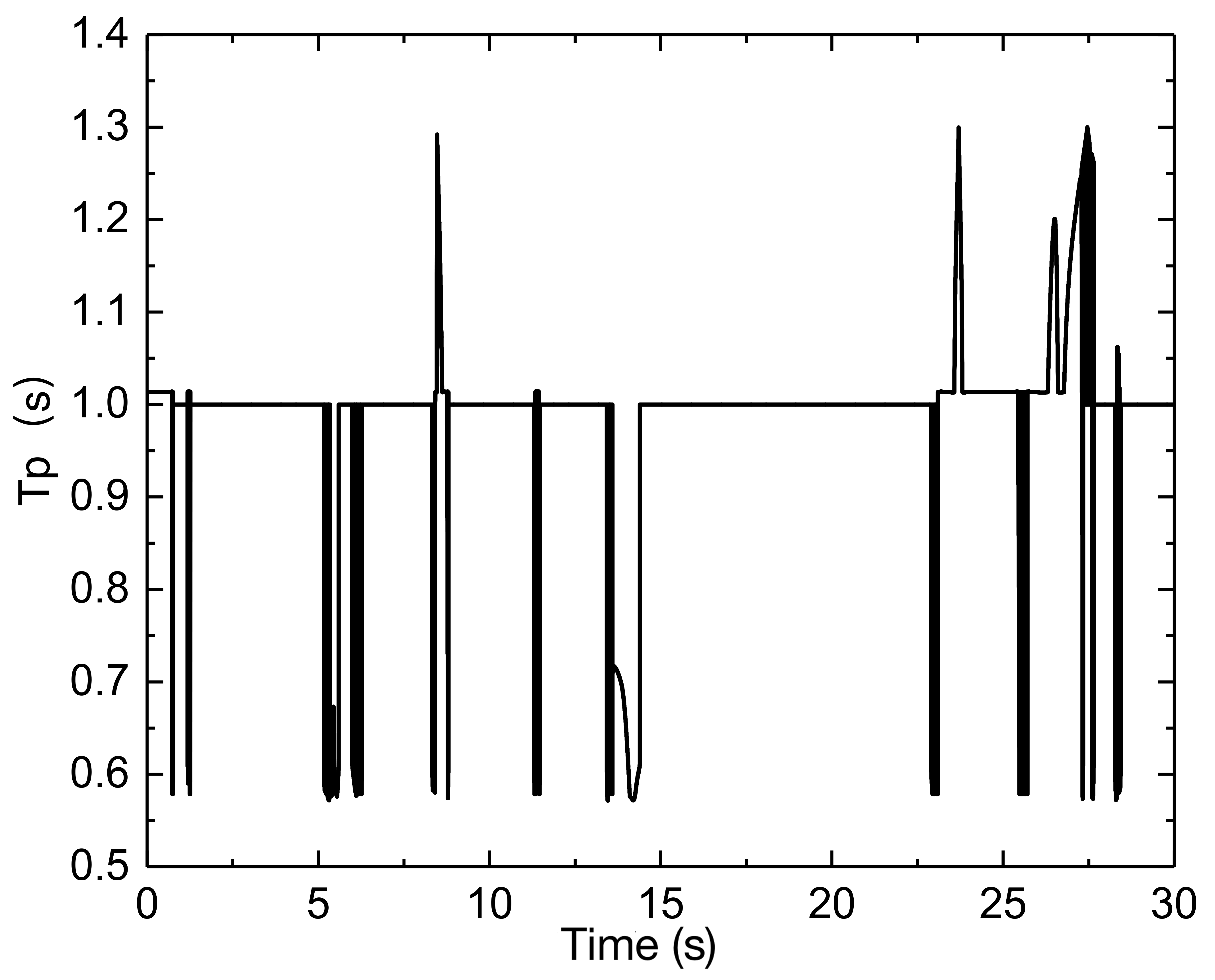}}
    \label{1a}\hfill
    \subfloat[ ]{
    \includegraphics[width=0.31\linewidth]{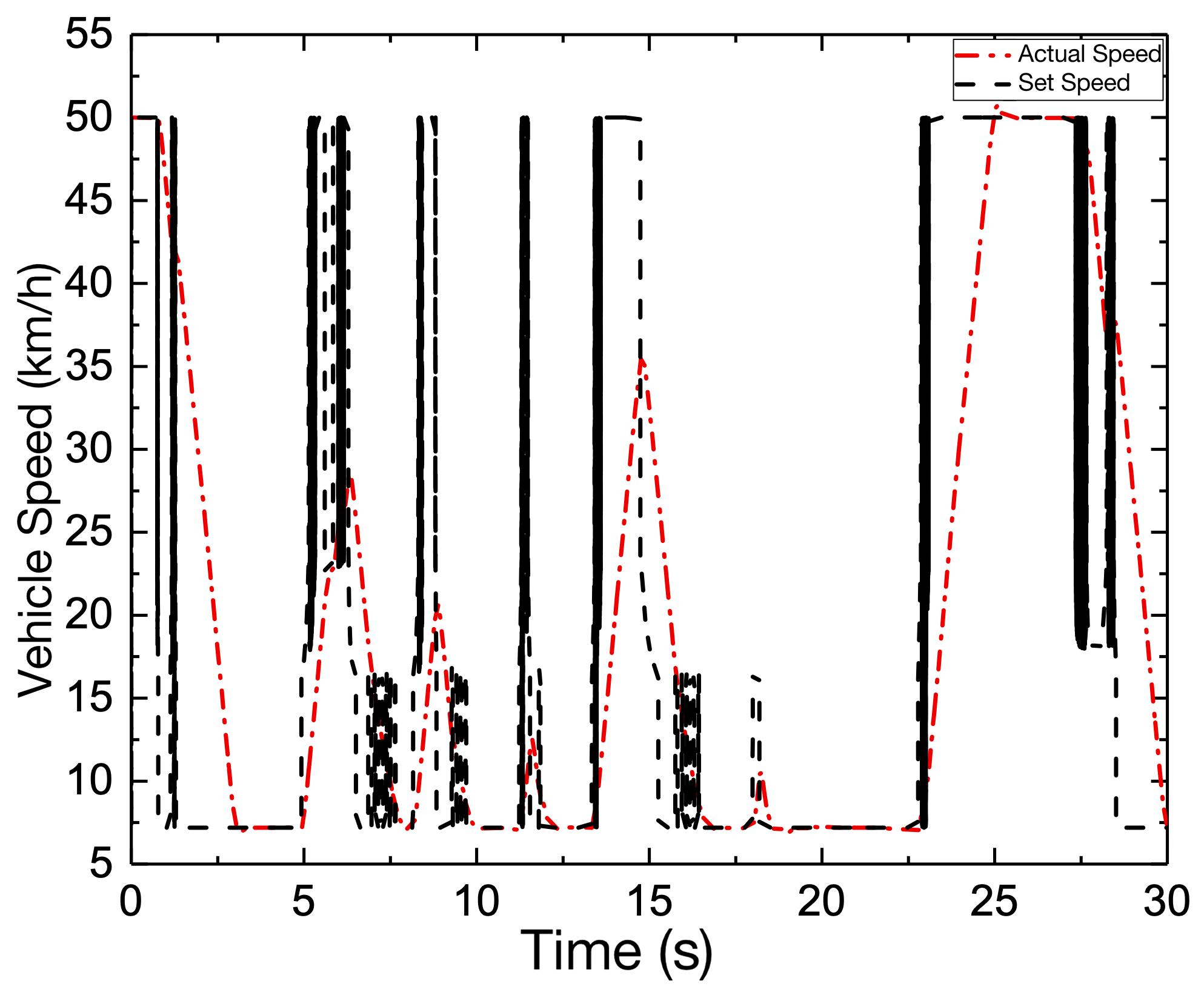}}
    \label{1b}\hfill
    \subfloat[ ]{
    \includegraphics[width=0.33\linewidth]{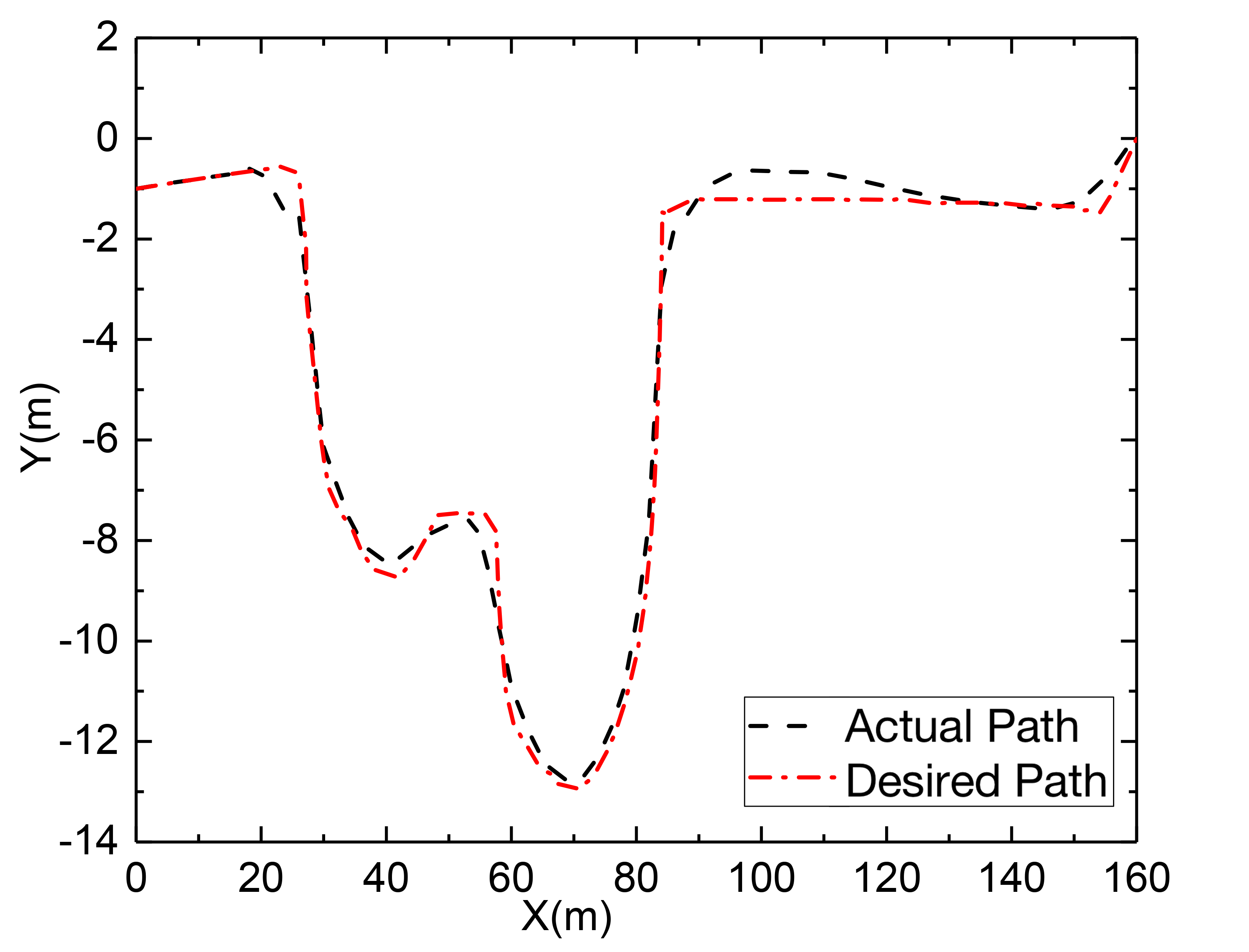}}
    \label{1c}\hfill
    \subfloat[ ]{
    \includegraphics[width=0.31\linewidth]{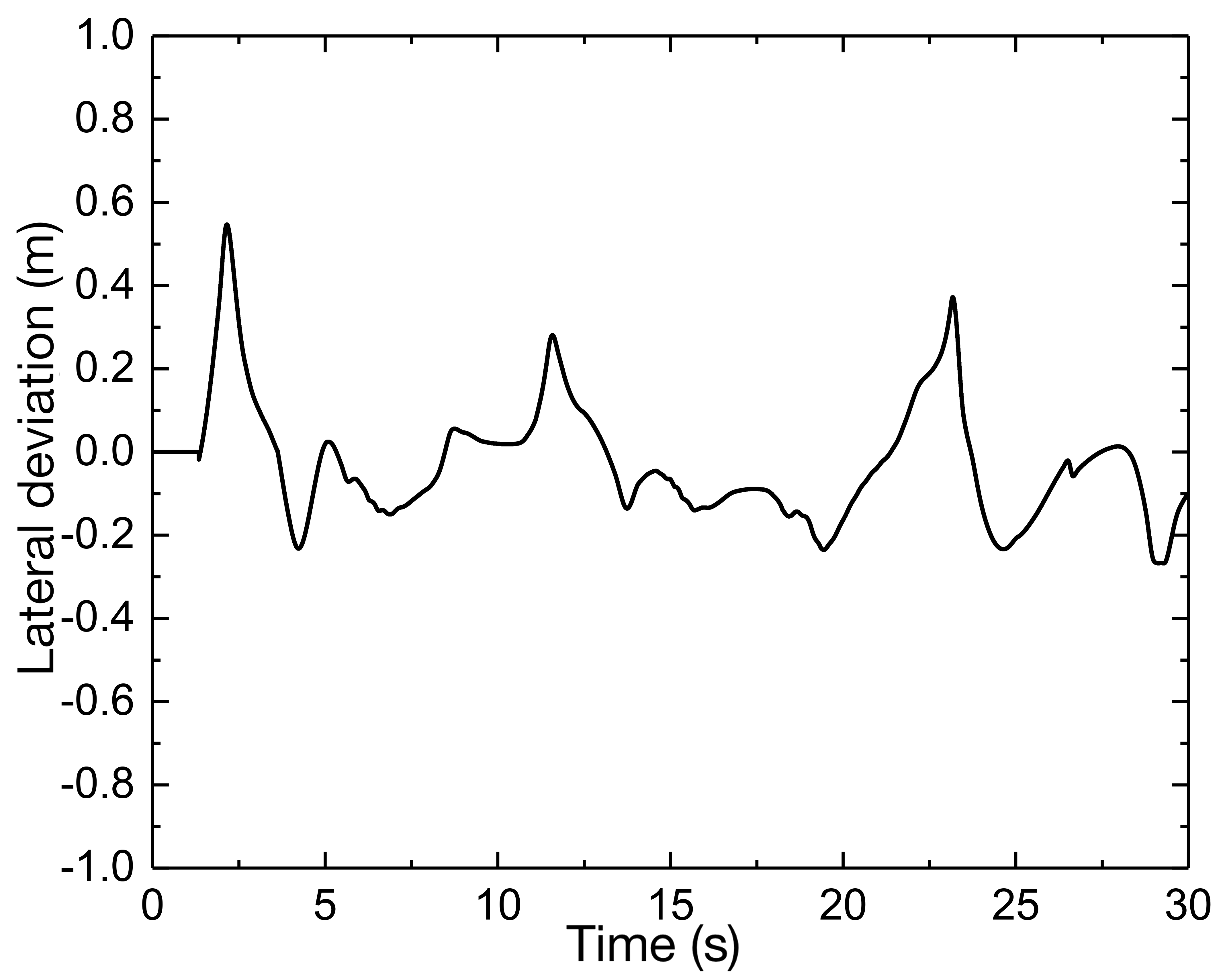}}
    \label{1d}\hfill
    \subfloat[ ]{
    \includegraphics[width=0.33\linewidth]{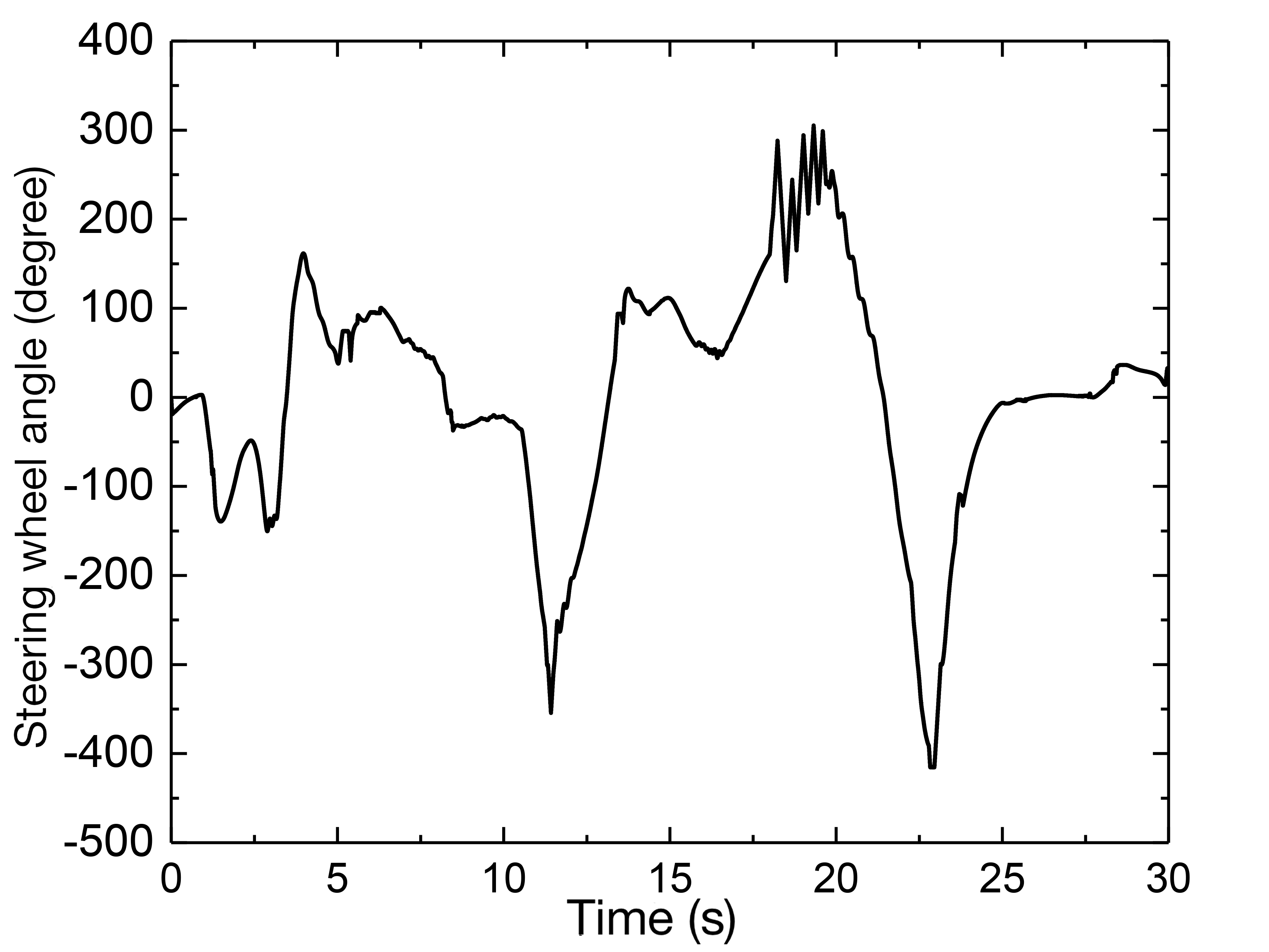}}
    \label{1e}\hfill
    \subfloat[ ]{
    \includegraphics[width=0.32\linewidth]{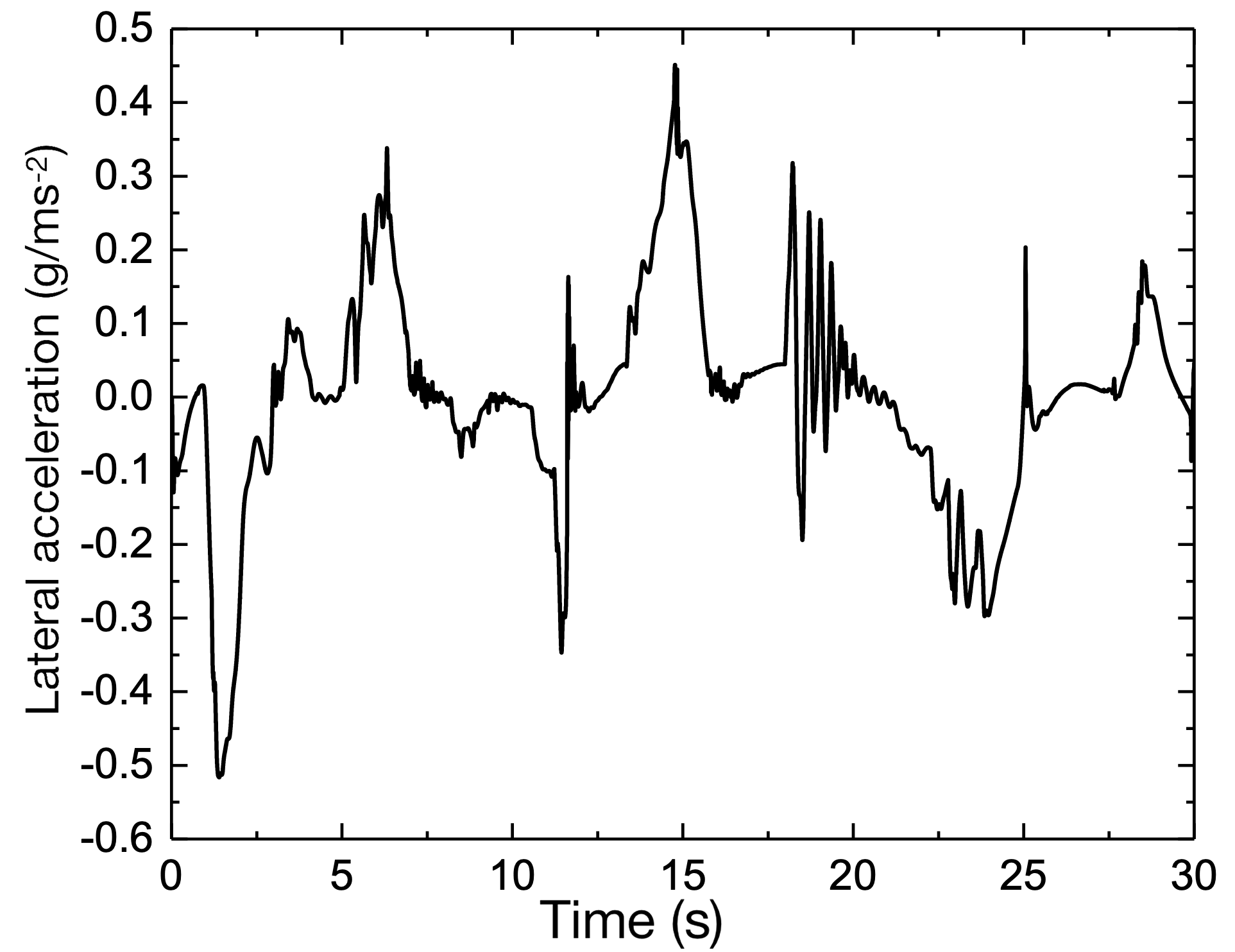}}
    \label{1f}\hfill
    \caption{(a) $T_{p}$ change curve; (b) Speed tracking results; (c) Path tracking results; (d) Lateral deviation results; (e) Corresponding steering wheel angle results; (f) Lateral acceleration results. The setting with CC-based method under 50 km/h.}
    \label{fig8}
\end{figure*}

Fig. \ref{fig8} (a) depicts the obtained preview time $T_{p}$ when the vehicle speed increases to 50 km/h, where the road curvature is large, the preview time and distance are both reduced, and the majority of the preview time is one second. Because the maximum adhesion given by the tires is regarded as the major speed restriction and the vehicle's limit in speed planning, when the speed is raised, however, the vehicle speed cannot be instantly lowered if the curvature is larger than 0.1. Therefore, it is necessary to slow down the vehicle earlier by establishing a lower speed threshold. The vehicle's speed planning and tracking results are illustrated in Fig. \ref{fig8} (b). The tracking path and accompanying lateral deviation are determined by lowering the vehicle speed immediately before approaching the curve and allowing the vehicle to finish the steering operation for varied curvatures, as shown in Figs. \ref{fig8} (c) and \ref{fig8} (d). It creates a greater deviation than 30 km/h since increasing the speed makes the vehicle more prone to side slip during the steering, and even if the vehicle is in a stable zone in advance, by reducing the speed, the actuators will have a delay in taking action. As a result, although the curvature calculation (CC-based) method is straightforward to be calculated, it does not take into account different restrictions and real-time feedback correction according to the vehicle's condition. Therefore, it will generate larger lateral deviations at higher vehicle speeds when the road curvature is rather considerable.

 Comparing the tracking results under different speeds, the CC-based tracking controller performs better when the vehicle speed is relatively lower in such a complicated environment. However, when the vehicle speed increases, it is not enough to reduce the speed to a relatively low level in advance where the curvature is significant. This is mostly because this method does not have real-time feedback correction of the vehicle's state and considers different limits. As a result, it is not ideally suitable for an environment with greater operational difficulty. On the other hand, this approach is straightforward and does not require significant computation, so it is more applicable under typical working conditions.

\subsection{Results of tracking control algorithm based on MPC}
 The simulation settings of vehicle speed 30 km/h and 50 km/h have been specified separately to validate the effectiveness of the tracking controller based on the MPC, with the control goal and the process of continuous feedback optimization.

(1) Vehicle speed is 30 km/h

Based on the established vehicle model with both lateral and yaw directions in MPC, the road surface adhesion coefficient is $\mu = 0.85$, and the MPC controller is written using the S-function in Simulink with parameters: sample period $T = 0.05s$, prediction step $N_p = 25$, control step $N_c = 10$, and the objective function weights are set as follows:
\begin{equation}
Q=\begin{bmatrix}
    &2000 &0 &0 &0\\ &0 &1000 &0 &0\\&0 &0 &1000 &0\\ &0 &0 &0 &1000
\end{bmatrix},R = 1.5 \times 10^5
\label{eq}   
\end{equation}

\begin{figure*}
    \centering
    \subfloat[ ]{
    \includegraphics[width=0.32\linewidth]{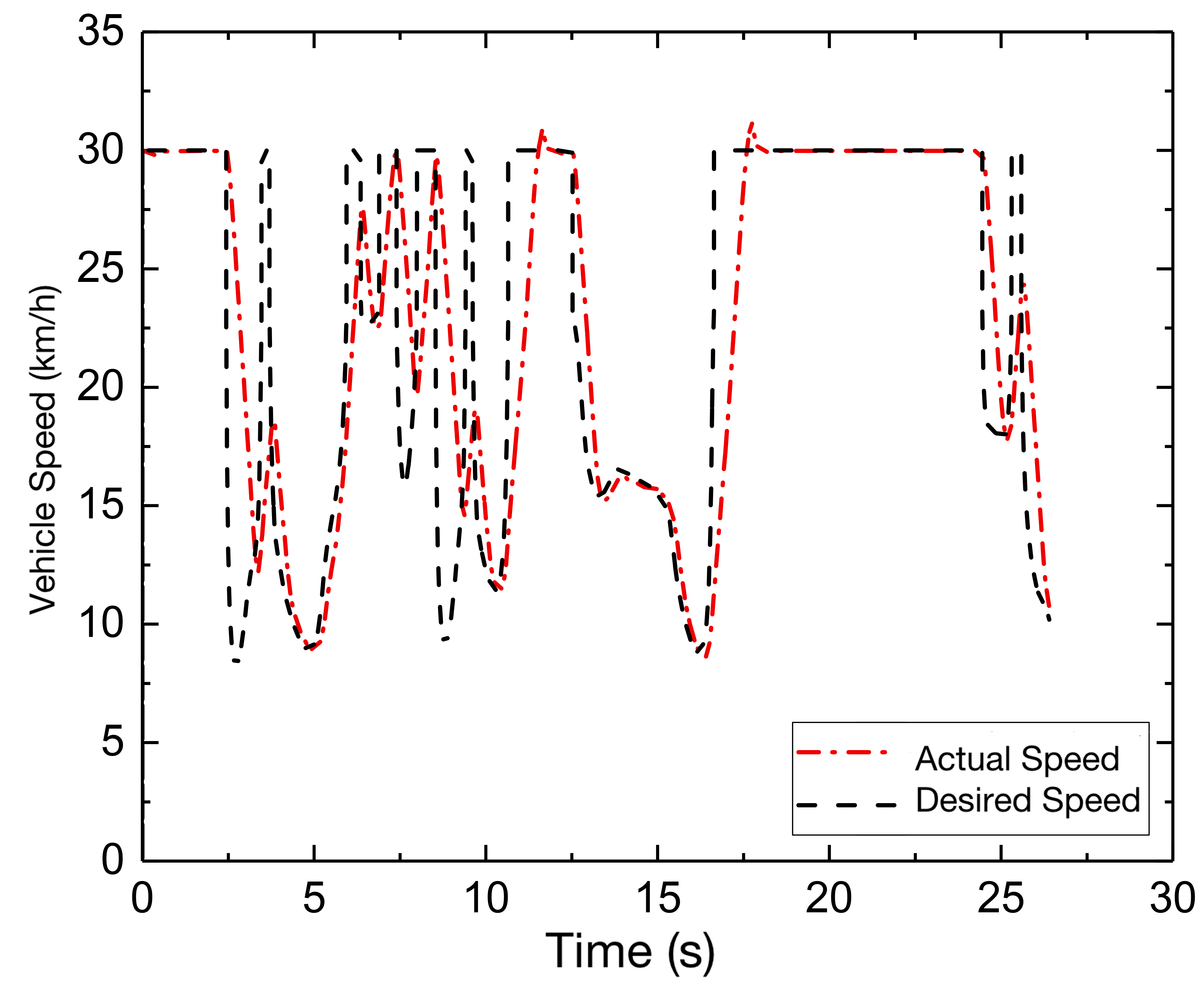}}
    \label{1a}\hfill
    \subfloat[ ]{
    \includegraphics[width=0.33\linewidth]{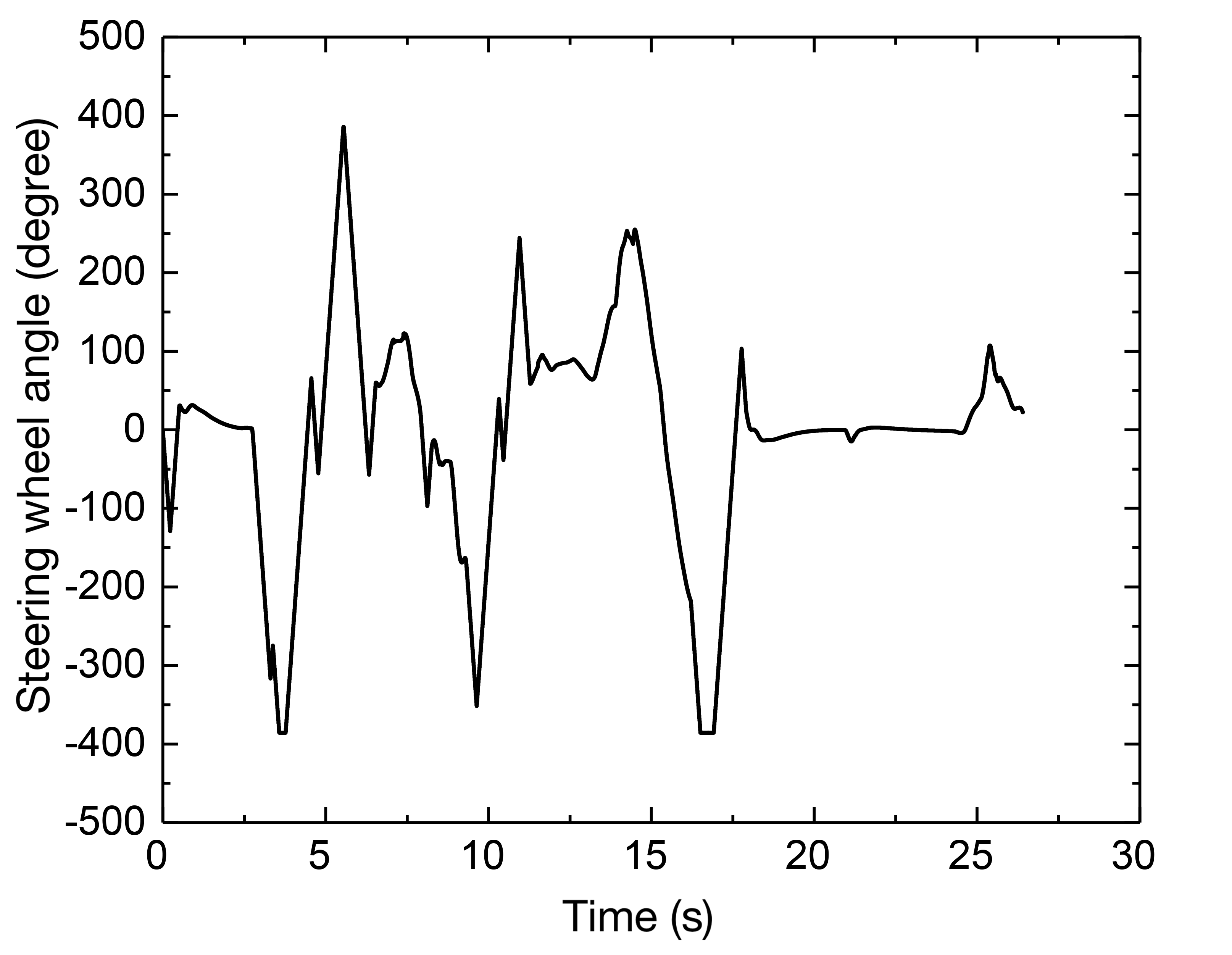}}
    \label{1b}\hfill
    \subfloat[ ]{
    \includegraphics[width=0.32\linewidth]{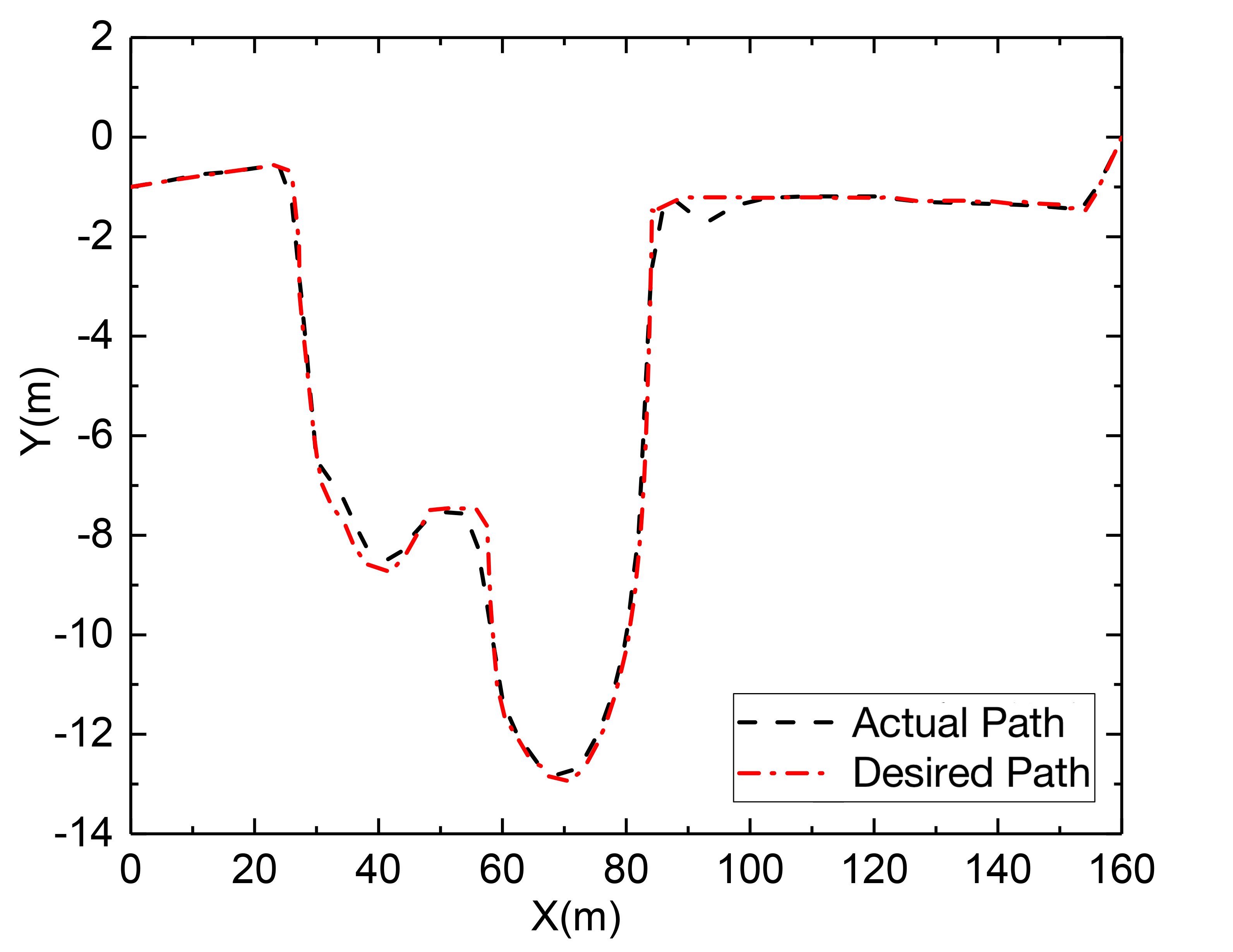}}
    \label{1c}\hfill
    \subfloat[ ]{
    \includegraphics[width=0.31\linewidth]{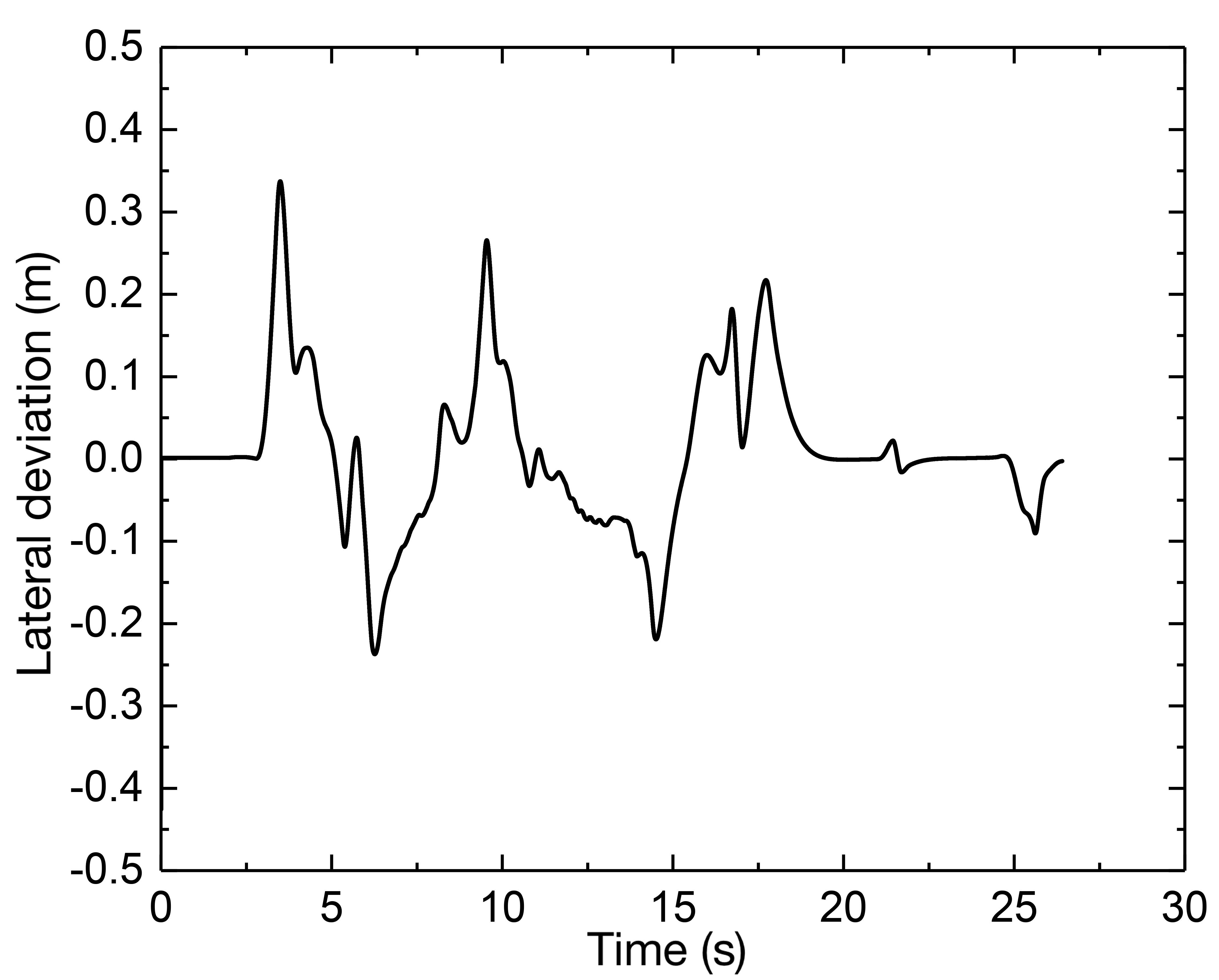}}
    \label{1d}\hfill
    \subfloat[ ]{
    \includegraphics[width=0.31\linewidth]{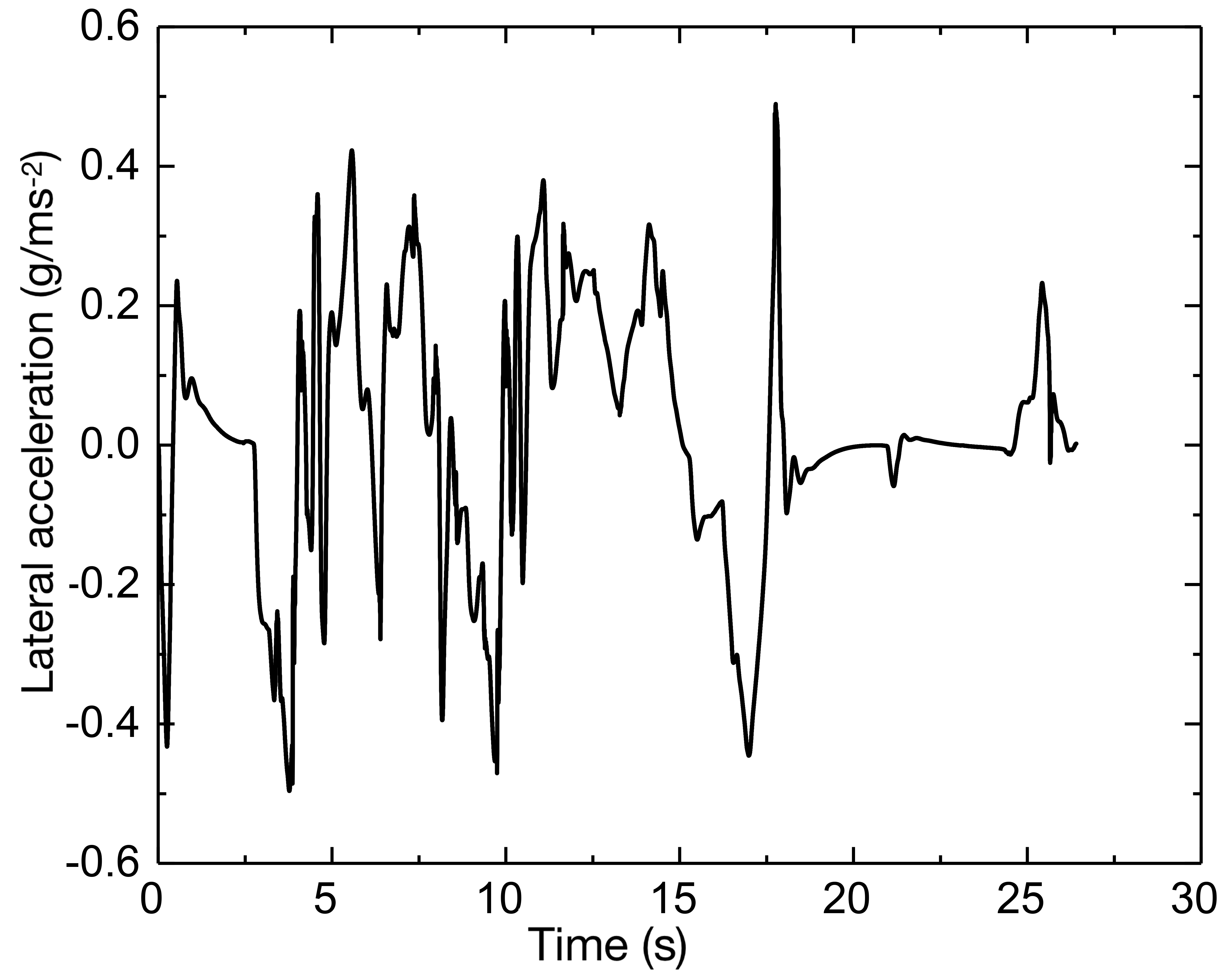}}
    \label{1e}\hfill
    \subfloat[ ]{
    \includegraphics[width=0.31\linewidth]{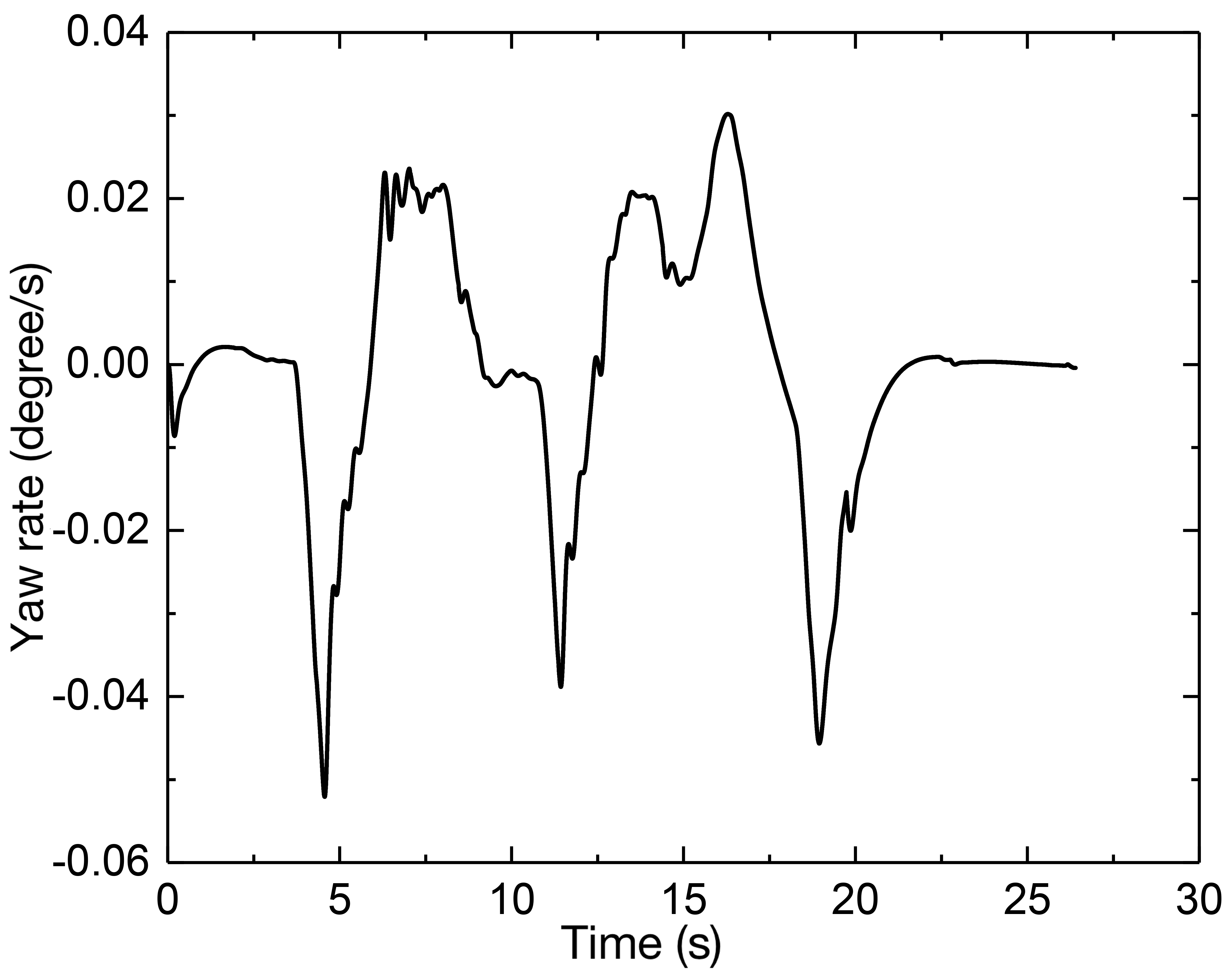}}
    \label{1f}\hfill
    \caption{(a) $T_{p}$ change curve; (b) Speed tracking results; (c) Path tracking results; (d) Lateral deviation results; (e) Corresponding steering wheel angle results; (f) Lateral acceleration results. The setting with MPC-based method under 30 km/h.}
    \label{fig9}
\end{figure*}

The slack factor is taken as 1000, the output deviation constraint is [-0.5, 0.5], and the front wheel steering angle constraint is [$-20^{\circ}$, $20^{\circ}$]. Therefore, the results can be obtained and presented in Fig. \ref{fig9}. In  Fig. \ref{fig9} (a), the vehicle can maintain the longitudinal speed; the tracking control precision is within 0.3 m, which has a strong tracking capability in extremely complicated environments. From the lateral acceleration, the vehicle is in a rather stable area; only when the curvature is at its greatest extent will the lateral acceleration reach its greatest value. This is because there will be a greater requirement to increase the lateral force.

(2) Vehicle speed is 50 km/h

\begin{figure*}
    \centering
    \subfloat[ ]{
    \includegraphics[width=0.32\linewidth]{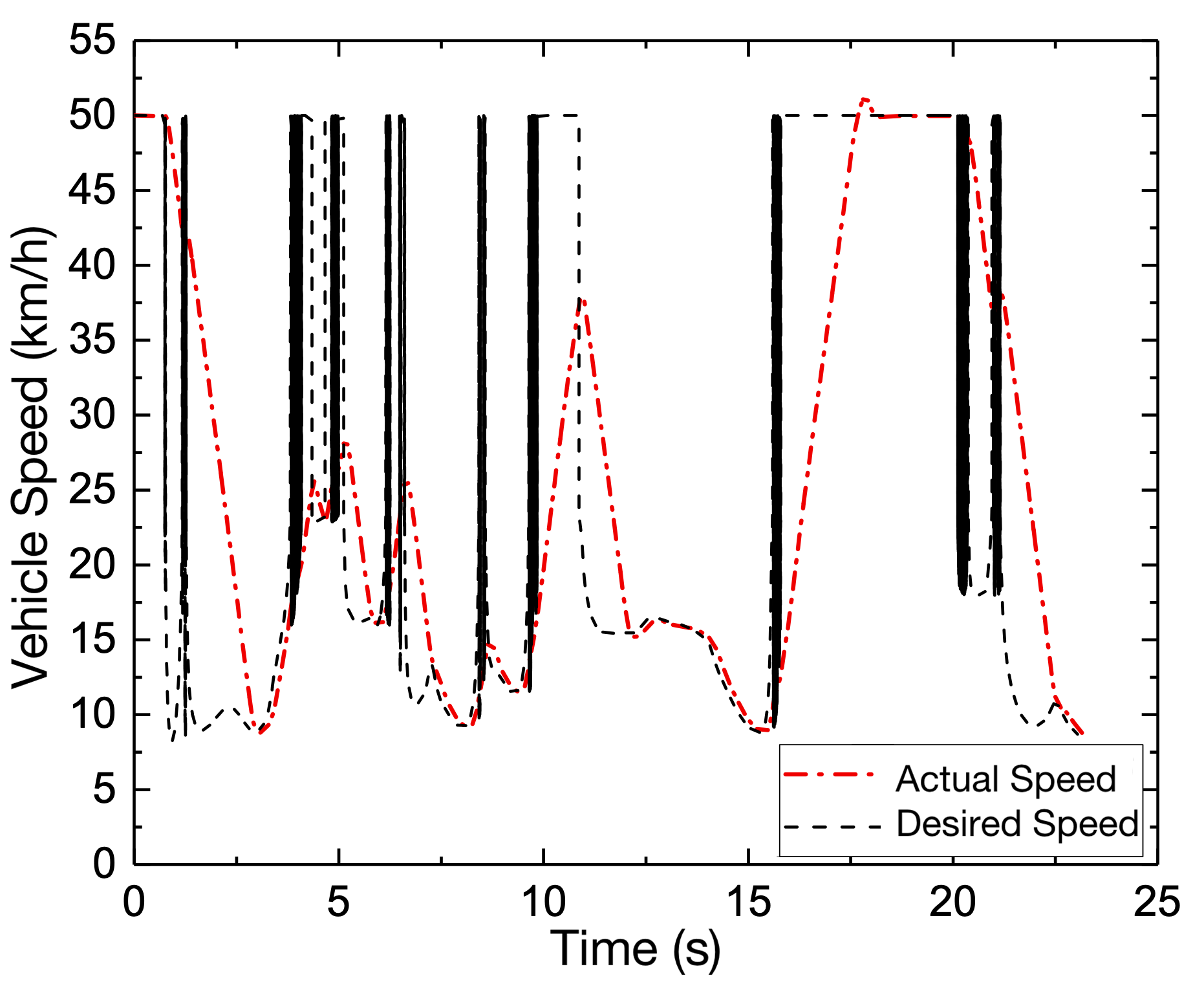}}
    \label{1a}\hfill
    \subfloat[ ]{
    \includegraphics[width=0.33\linewidth]{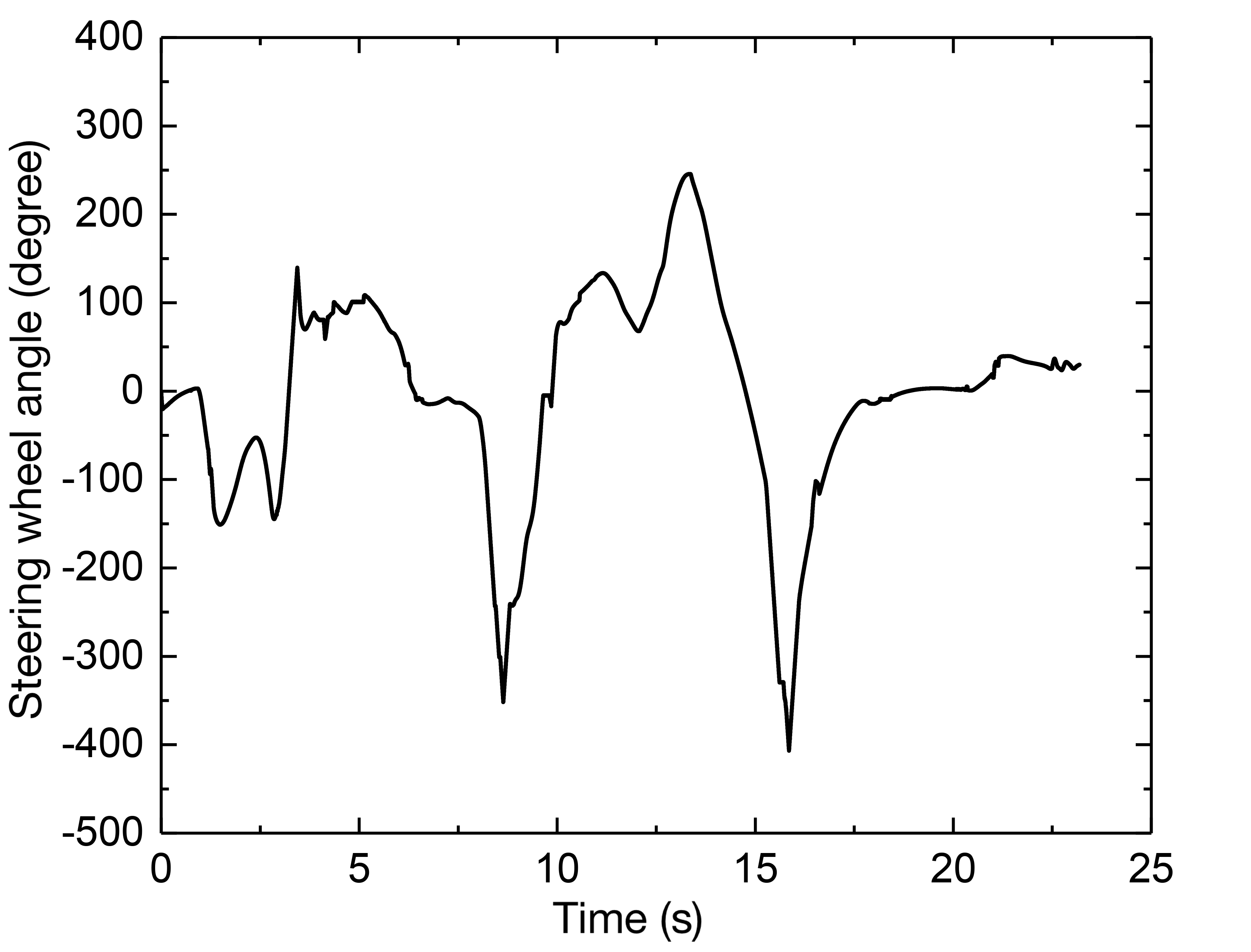}}
    \label{1b}\hfill
    \subfloat[ ]{
    \includegraphics[width=0.32\linewidth]{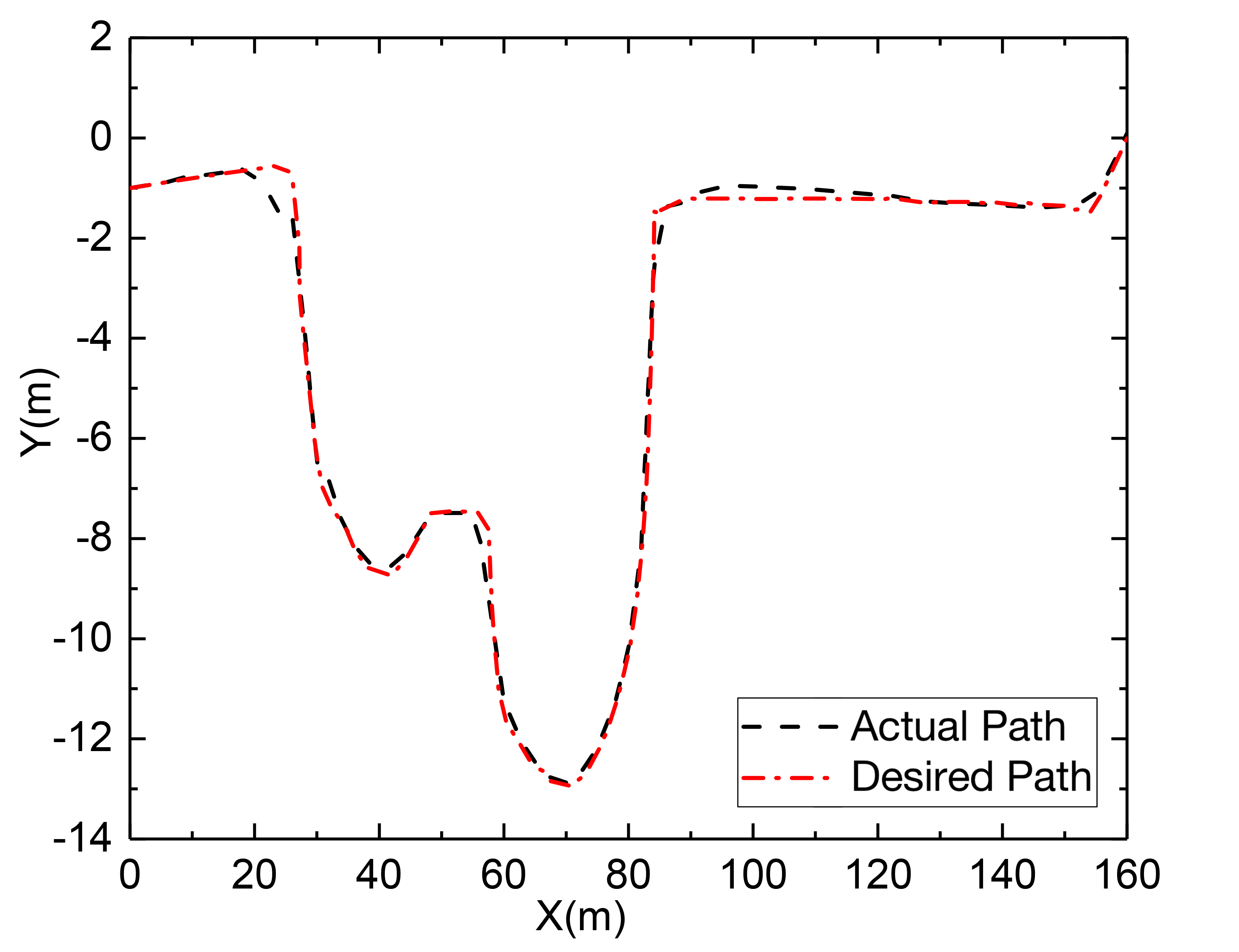}}
    \label{1c}\hfill
    \subfloat[ ]{
    \includegraphics[width=0.32\linewidth]{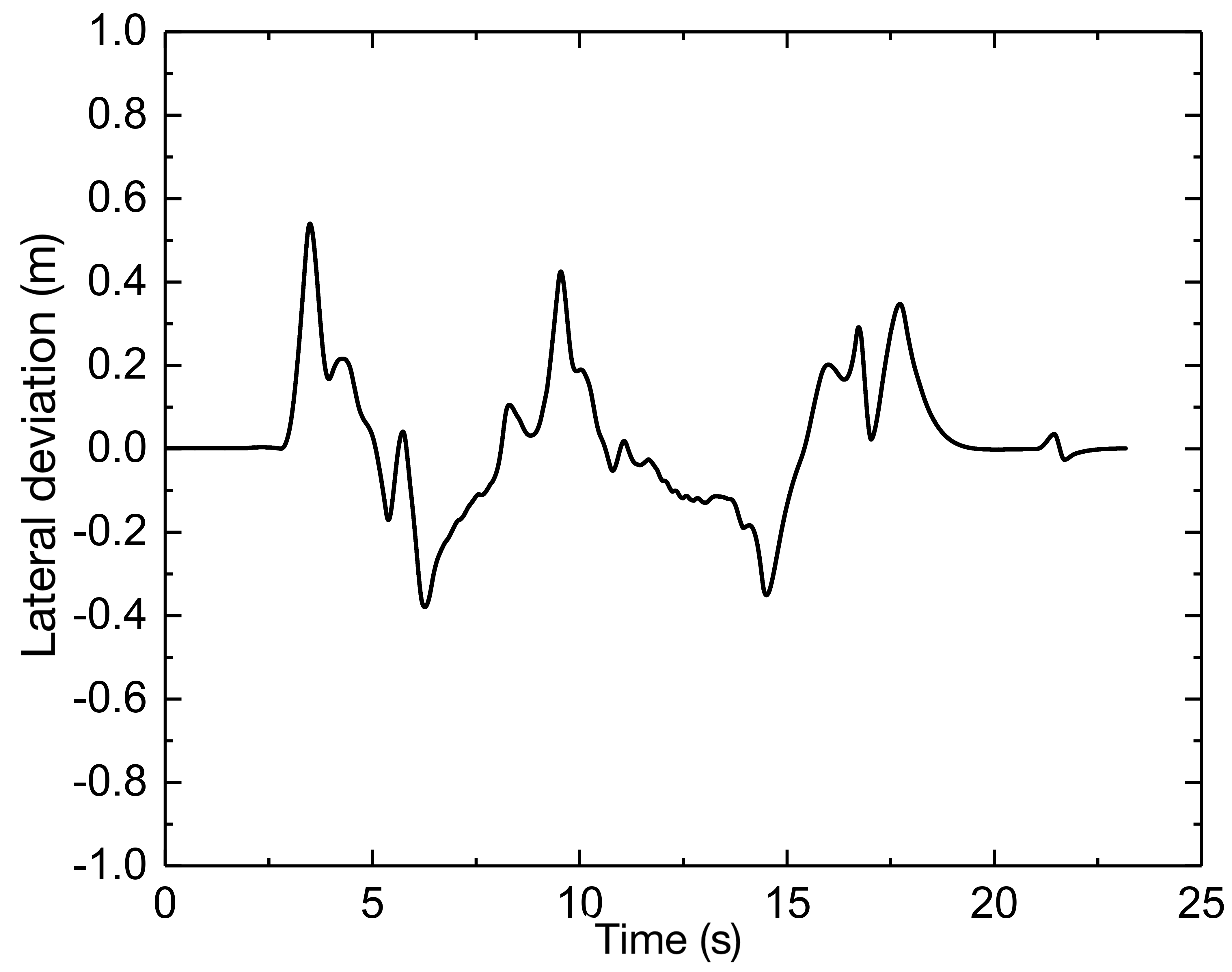}}
    \label{1d}\hfill
    \subfloat[ ]{
    \includegraphics[width=0.32\linewidth]{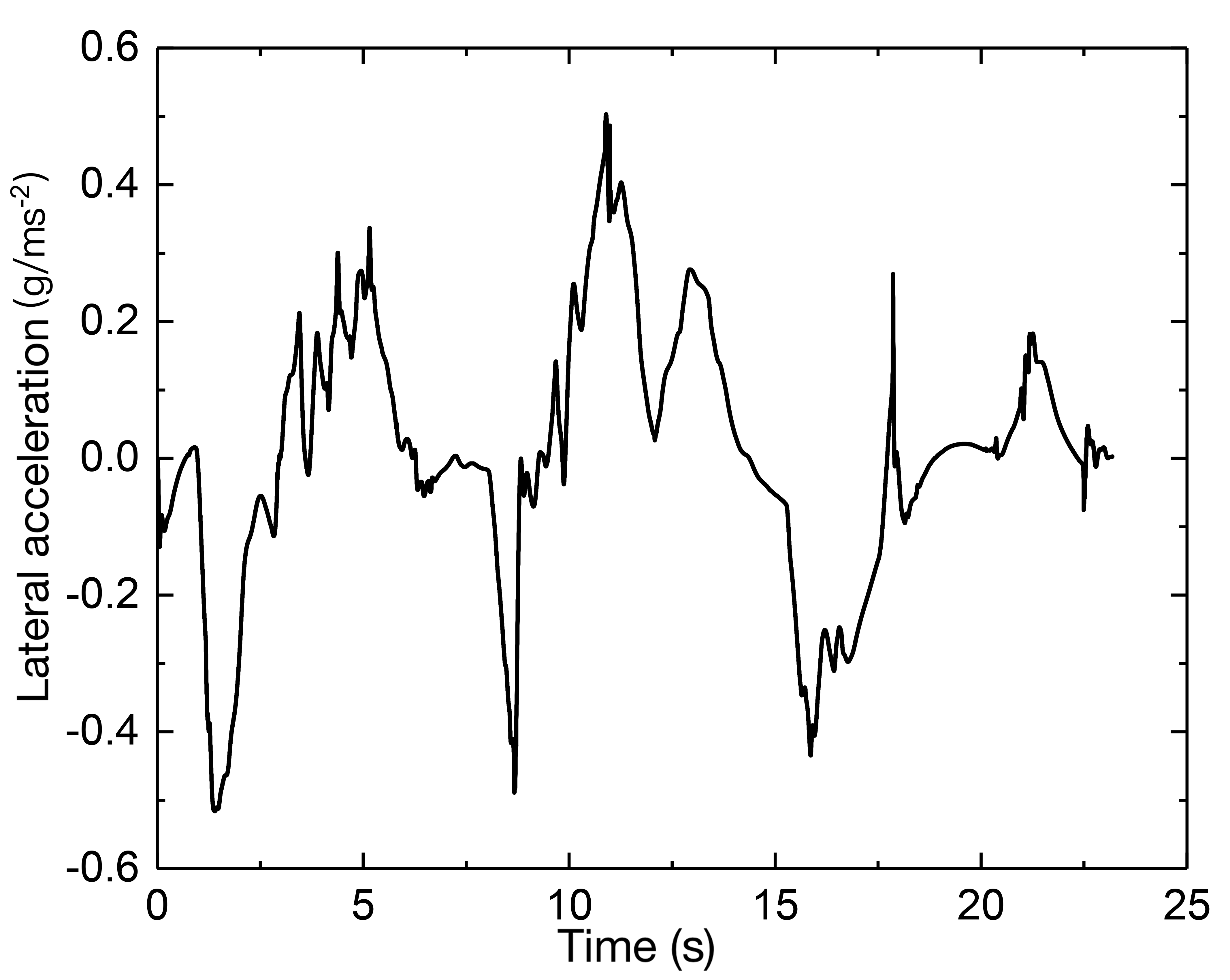}}
    \label{1e}\hfill
    \subfloat[ ]{
    \includegraphics[width=0.32\linewidth]{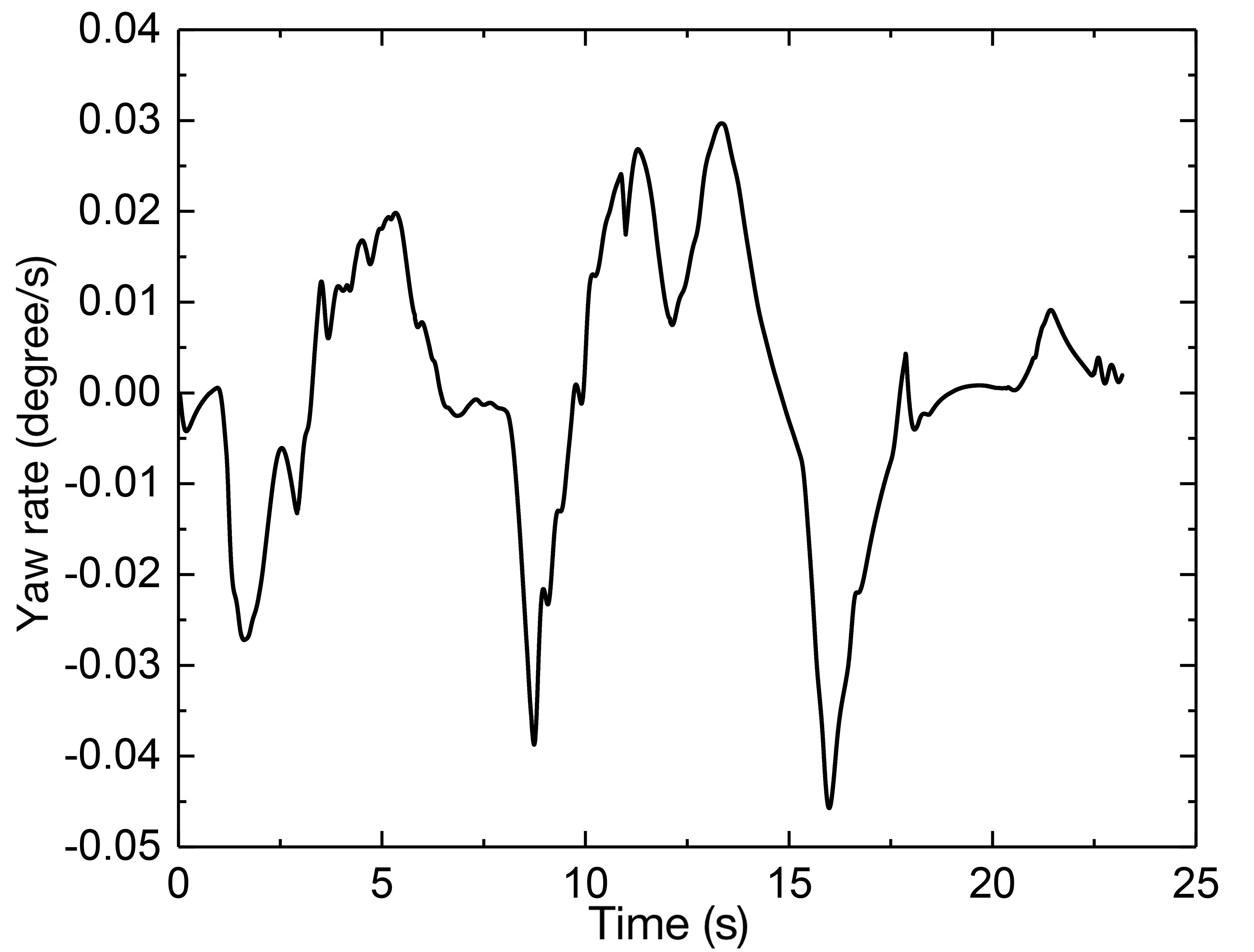}}
    \label{1f}\hfill
    \caption{(a) $T_{p}$ change curve; (b) Speed tracking results; (c) Path tracking results; (d) Lateral deviation results; (e) Corresponding steering wheel angle results; (f) Lateral acceleration results. The setting with MPC-based method under 50 km/h.}
    \label{fig10}
\end{figure*}

The simulation results are depicted in Fig. \ref{fig10} following an increase in the vehicle speed. The vehicle is capable of maintaining the desired speed. The path control deviation in this complicated environment can be controlled to within about 0.5 m, and it can be seen from the lateral acceleration and yaw rate that an increase in vehicle speed does not result in a decrease in the performance of the vehicle's stability. Similarly, except for the abrupt bends with the greatest curvature, the lateral acceleration will reach its maximum to give an adequate amount of lateral force, but the longitudinal speed will decrease. Because the combined effect is required to help the vehicle safely in a more complicated environment, in which decoupling the longitudinal and lateral motion is tough.

MPC tracking controller can produce satisfactory tracking results as well, when the speeds are relatively modest, there is no significant difference in the tracking precision provided by the two methods. However, in a complicated environment,  when the vehicle speed rises, the curvature is still rather significant in some locations. MPC with the vehicle dynamics model can adjust the vehicle's state and predict the vehicle's output for a future period by continuous feedback optimization correction. Consequently, it also has an improved tracking capacity at faster speeds.

\section{Conclusion}
Based on the comprehensive control strategy of FDWEV chassis, different tracking controllers are designed to realize the trajectory tracking control of the vehicle in low and medium-speed working conditions with considerable curvatures to verify the presented algorithms. First, the repulsive force function and the direction are optimized to solve the unreachability of the target point in the standard APF approach. Then the corresponding speed planning is designed for the kinematic and dynamic constraints of the vehicle. Along with the kinematic and lateral dynamics models, two tracking controllers based on curvature calculation and MPC are designed in a complicated unstructured environment to improve tracking accuracy and stability. Furthermore, co-simulation in Simulink/Carsim/Amesim was conducted to validate the presented algorithms. Overall, based on the relatively significant curvature of the generated path, the two tracking controllers have strong tracking capabilities under low-speed working conditions with about 0.3 m deviation; when speed increases, the tracking capabilities can reduce, and the MPC-based controller with 0.5 m deviations can improve the following ability while maintaining better control compared to the CC-based controller with 0.6 m deviation. Hence, MPC can handle the optimization problem with constraints better in more dynamic unstructured environments.

Future studies will be focused on the real-world comparison of the two methods and the integration of collision avoidance control with dynamic environment information by radar and vision sensing systems.

\bibliographystyle{IEEEtran}
\bibliography{Refs}



\end{document}